\documentclass[3p,final,11pt,authoryear]{elsarticle}
\usepackage{subcaption}
\usepackage{rotating}
\usepackage{tikz}
\usepackage{booktabs}
\usepackage{graphicx}
\usepackage{amssymb}
\usepackage{amsthm}
\usepackage{multirow}
\usepackage{amsmath}
\usepackage[linesnumbered,ruled]{algorithm2e}
\newsavebox{\measurebox}

\usepackage{lineno}
\usepackage{array,calc}
\usepackage{lscape}
\usepackage{float}
\usepackage{hyperref}
\usepackage[nameinlink,capitalize]{cleveref}

\floatstyle{plaintop}
\restylefloat{table}

\usepackage{xcolor}
\usepackage{adjustbox}
\usepackage{tabularx}
\usepackage[symbol,multiple]{footmisc}
\usepackage{array,calc}
\usepackage{makecell}

\newcommand\arash[1]{\textcolor{black}{#1}}

\journal{Transportation Research Part C}

\begin{document}
\begin{frontmatter}
\title{A context-aware pedestrian trajectory prediction framework for automated vehicles}
\author[1]{Arash Kalatian\footnote{Corresponding author}}
\author[2]{Bilal Farooq}

\address[1]{Institute for Transport Studies, University of Leeds, UK} 
\address[2]{Laboratory of Innovations in Transportation (LiTrans),

 Ryerson University, Toronto, Canada}

\begin{abstract}
With the unprecedented shift towards automated urban environments in recent years, a new paradigm is required to study pedestrian behaviour. Studying pedestrian behaviour in futuristic scenarios requires modern data sources that consider both the Automated Vehicle (AV) and pedestrian perspectives. Current open datasets on AVs predominantly fail to account for the latter, as they do not include an adequate number of events and associated details that involve pedestrian and vehicle interactions. To address this issue, we propose using Virtual Reality (VR) data as a complementary resource to current datasets, which can be designed to measure pedestrian behaviour under specific conditions. In this research, we focus on the context-aware pedestrian trajectory prediction framework for automated vehicles at mid-block unsignalized crossings. For this purpose, we develop a novel multi-input network of Long Short-Term Memory (LSTM) and fully connected dense layers. In addition to past trajectories, the proposed framework incorporates pedestrian head orientations and distance to the upcoming vehicles as sequential input data. By merging the sequential data with contextual information of the environment, we train a model to predict the future pedestrian trajectory. Our results show that the prediction error is reduced by considering contextual information extracted from the crossing environment, as well as the addition of time-series behavioural information to the model. To analyze the application of the methods to real AV data, the proposed framework is trained and applied to pedestrian trajectories extracted from an open-access video dataset. Finally, by implementing a game theory-based model interpretability method, we provide detailed insights and propose recommendations to improve the current automated vehicle sensing systems from a pedestrian-oriented point of view. 
\end{abstract}

\end{frontmatter}
\section{Introduction}
\label{S:TInt}

The rapid technological development in Automated Vehicles (AVs), followed by a tremendous increase in their adoption, promises a significant transformation in the dynamics of urban roads. An important transformation expected to occur is the effect of AVs on pedestrians, as the most vulnerable road users. Particularly, the absence of a driver in the vehicle leads to the absence of eye contact and observation of head and body movements by the driver. Therefore, there is a strong need to re-examine the interaction between vehicle and pedestrian, while accounting for the expected changes. To be able to compensate for the silent agreement currently between the driver and the pedestrian and establish a similar type of interactions between them in an automated environment, AVs need to find a way to anticipate pedestrian behaviour, i.e. intentions, choices and movements/trajectories, based on the pedestrian reactions and postures the AV captures. Failures in predicting pedestrian behaviour and the absence of timely actions by the AV have already resulted in catastrophic accidents in recent years, even at very slow speeds~\citep{uber,vienna}. Studying pedestrian behaviour is an active and extensive area of research. However, we focus on pedestrian behaviour when crossing mid-block, unsignalized roads. As rule-obeying AVs find their way on the streets in the future urban spaces, it is a likely scenario that the proportion of this type of crossing increases~\citep{millard2018pedestrians}. On the other hand, by going through the official reports of Uber's test AV incidence in Arizona, it can be concluded that the vehicle's sensing system could not predict the pedestrian's path correctly because she was crossing mid-block, and \textit{``the system design did not include consideration for jaywalking pedestrians.''}~\citep{uber,ntsb}. Thus, a thorough investigation of mid-block crossings is timely and of vital importance. 

\begin{figure}
    \centering
    \includegraphics[scale=0.025]{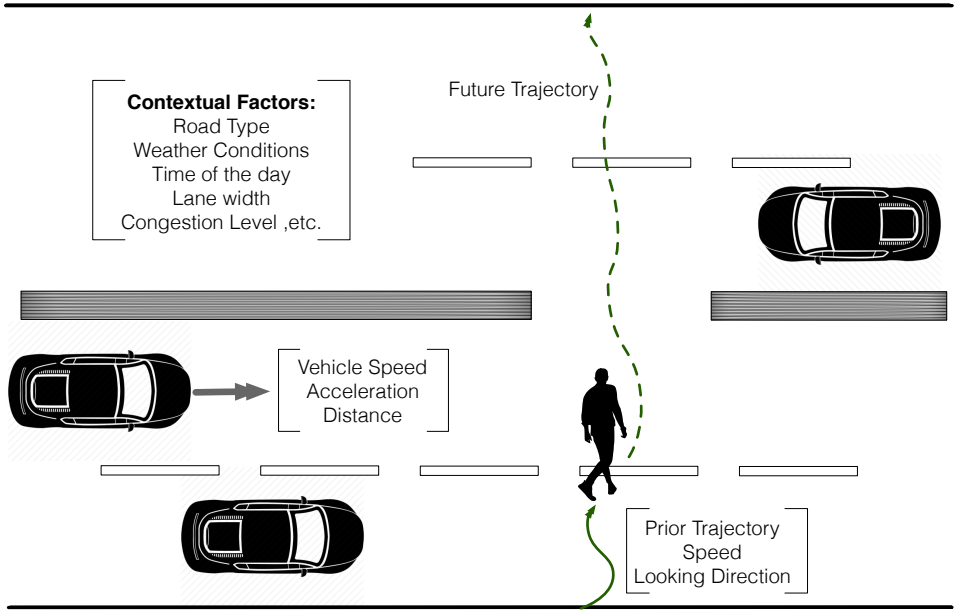}
    \caption{Illustration of possible involving factors in pedestrian trajectory prediction}
    \label{fig:Tinteraction}
\end{figure}

At a conceptual level, we can simplify the interactions of an AV and a pedestrian crossing mid-block to three parts~\citep{kalatian2020decoding}: (a) a pedestrian waits on the sidewalk for the right time to initiate a cross, (b) he/she follows a certain trajectory based on the characteristics of the approaching vehicle and geometric and environmental conditions, (c) the approaching vehicle anticipates pedestrian behaviour and reacts by making the required decisions to provide a safe and comfortable interaction for both the pedestrian and the passengers. We explored the first part of this interaction, i.e., wait time of a pedestrian, in our two previous studies~\citep{kalatian2019deepwait,kalatian2020decoding} and others have optimized AV behaviour based on approaching pedestrians~\citep{vasquez2019multi}. In this study, the focus is to understand and develop prediction models for the second part, i.e. the pedestrian trajectory. As depicted in \cref{fig:Tinteraction}, various factors might contribute to the trajectory followed by a pedestrian while crossing the street. Prior actions by the pedestrian and vehicle, as well as the features of the environment in which the cross takes place, can be used to predict the next movements of the pedestrian. Together, both part (a) and part (b) predictive models can be integrated and utilized by AVs to understand and predict pedestrian behaviour more accurately and to proactively make maneuvering decisions~\citep{vasquez2019multi}.

In this study, we first provide an extensive review of the open-access AV datasets from a pedestrian-oriented point of view and discuss existing gaps within them. \arash{As all the currently available open-access AV datasets fail to provide an adequate number of mid-block crossing events and rich contextual information, we then propose using Virtual Reality (VR) controlled experiments} as a complementary tool to better understand the pedestrian behaviour under specific conditions. A novel multi-input network of Long Short-Term Memory (LSTM) and fully connected dense layers is developed to model pedestrian trajectory while crossing a road in an automated environment. In the proposed model, time-series data of the initial steps of crossing are added to non-time-series data of contextual information of the crossing's environment to predict the next steps of pedestrian trajectories. The proposed framework is first trained and tested on VR data, and then on an open-access video dataset to analyze the applicability of the framework to real datasets.  A game theory-based post-hoc interpretability method for neural networks is then applied to analyze the contributing factors to pedestrian trajectory prediction accuracy. By providing insights into the most important factors in trajectory prediction, we propose suggestions that can improve currently available datasets from a pedestrian-oriented point of view. \arash{This study contributes to the transportation research community by proposing and utilizing virtual reality data as a complementary tool for pedestrian trajectory data collection in the context of automated vehicles. Providing an extensive and pedestrian-oriented review of the currently available AV datasets and presenting insights on their drawbacks and ways to address these drawbacks is another contribution of this research study, which we believe has not been adequately addressed in the literature. }

The rest of this paper is organized as follows: a review of relevant studies in trajectory prediction and an extensive review of currently available open-access AV datasets are provided in the next section. \cref{S:t3} briefly discusses data collection and pre-processing procedures. Methodology and proposed architecture are described in \cref{S:t4}. The application of our proposed framework on the data and their interpretation are discussed in \cref{S:t5}. Finally, \cref{S:t6} is dedicated to conclusions, final remarks, and future research plans.
\section{Background}
\label{S:tBack}
This section provides an overview of the research studies on pedestrian crossing behaviour, emphasizing the trajectory prediction studies. Traditional approaches to pedestrian trajectory modelling and recent data-driven trends are discussed in this chapter. Modern data-driven approaches require novel large-scale datasets. In order to understand pedestrian behaviour in the presence of AVs, real datasets from AV manufacturers are the most reliable resources. Thus, a review of the available datasets from a pedestrian-oriented point of view is provided in this section.   
\subsection{Pedestrian Trajectory Prediction}
Early models in the literature tried to model the pedestrian movement using the concepts and theories of ideal gases~\citep{henderson1974fluid} or fluids~\citep{helbing1998fluid}. However, the turning point in pedestrian movement modelling was the social force model by \cite{helbing1995social}. Based on the idea that behavioural changes are caused by so-called social fields, Helbing and Molnar described forces affecting pedestrian behaviour as a result of the internal motivations of an individual to decide and perform actions. Researchers later calibrated social force models based on the purpose of their studies. In a study investigating pedestrian behaviour at signalized crosswalks, for instance, \cite{zeng2014application} incorporated forces from conflicting vehicles, signal phase and crosswalk boundaries to develop a modified social force model. The authors later added route plans, pedestrian acceleration or deceleration choice and their leader-follower behaviour as well as other underlying characteristics of pedestrians to further calibrate the social force model~\citep{zeng2017specification}. \cite{antonini2004simulation} applied random utility maximization based discrete choice models to pedestrian movement analysis using video data. The microscopic approach of the model allowed a detailed analysis of pedestrian movement. The choices that a pedestrian was facing at a certain time in their model were: (1) speed level and (2) discrete radial direction. Utility functions for each of these choices were defined based on the presence of obstacles, proximity to the destination and positions and speeds of other pedestrians. Later studies on these models added other variables, helping the model gain strength by observing various factors. For instance, \cite{guo2012route} added visibility parameters to the model while \cite{asano2010microscopic} later incorporated density. The major weaknesses of such microscopic methods are: 1. their highly myopic nature, as most of them focus on the immediate interactions and behaviours of pedestrians, and 2. their need of hand-craft functions, which makes it difficult to apply them to more complex settings~\citep{alahi2016social}. Moreover, models using logit formulation require the modeller to discretize the space and speed into arbitrary levels.

The widespread success of machine learning methods in recent years, as well as the availability of large pedestrian datasets, have resulted in a shift of pedestrian research trends to data-driven. In particular, recurrent networks, i.e., RNN and LSTM, have been the dominant machine learning methods for trajectory prediction. In most cases, the input data used for trajectory prediction is only the past trajectory of the pedestrians~\citep{xue2019location,zhang2019sr,alahi2016social,gupta2018social}. \cite{alahi2016social} introduced Social LSTM, a method that incorporated interactions among pedestrians in sequential models, namely Long Short-Term Memory (LSTM), to forecast pedestrian trajectory using video footage of walking individuals in crowded scenes. In their method, a \textit{Social Pooling Layer} is added to the framework. Through this layer, LSTM layers trained for individuals in a scene share their information. Despite the success of the Social-LSTM model in forecasting pedestrian trajectory, this model does not account for the contextual information on the environment and aspects like where the pedestrian is looking. The model may be applicable in pedestrian-dominant environments, e.g., shopping malls or train stations, but it is difficult to apply to situations like road crossing behaviours of pedestrians in an automated environment. Furthermore, the future trajectory predicted by Social-LSTM assumes a fixed-length future trajectory.  More recently, ~\cite{gupta2018social} introduced \textit{Social-GAN}, a model that predicts socially acceptable trajectories by training adversarial against a recurrent discriminator. Similar to~\citep{alahi2016social}, this model fails to capture context information from the environment, and its applicability is limited.

Some other research studies have incorporated contextual and semantic information into the pedestrian trajectory to predict their next locations~\citep{lee2017desire,chandra2019traphic,rasouli2019pie}. The semantic information used can be occupancy maps~\citep{bi2019joint}, road topology~\citep{casas2018intentnet}, the shape of objects~\citep{chandra2019traphic}, and the speed of the vehicles~\citep{bhattacharyya2018long,rasouli2019pie}. \cite{lee2017desire}, for instance, added semantic contexts, such as road structure and interactions and dynamics of the agents to their proposed RNN model and predicted pedestrian trajectory of variable lengths in a video dataset. Despite all the diversity in models, data and objective functions in the relevant research studies, only a few models have attempted to predict pedestrian trajectory in the context of AVs. In most such cases, analyzing pedestrians is limited to crowds and not in the context of interactions with vehicular traffic. To the best of our knowledge, almost all the models in the literature have used general-purpose video footage of pedestrian movements as the input data. Because pedestrian road crossing behaviour in the presence of AVs may differ, current datasets and scenarios may not apply to futuristic scenarios. One of the reasons behind this gap can be traced back to the lack of pedestrian data in open-access AV datasets. In the next subsection, an overview of these datasets is provided.

\subsection{Open-Access AV Datasets}
Despite the significant improvements in AV's general operations through video and sensor datasets, pedestrian-AV interaction modelling using such datasets has not been thoroughly explored. AV datasets have been widely explored for object detection~\citep{chang2019argoverse}, semantic segmentation~\citep{porzi2019seamless}, and vehicle trajectory prediction~\citep{gu2020lstm,chandra2020forecasting,lee2017desire}. However, a lack of pedestrian behavioural studies using these open-access AV datasets made us explore the gaps and missing components in these datasets. For the purpose of this study, we investigated four open-access AV datasets. These datasets were selected among several available datasets because of their proper annotations of pedestrians and vehicles over consecutive frames, the inclusion of urban areas and the duration of the drives.

NuScenes dataset was the first AV dataset investigated. With 1,000 scenes of 20 seconds each~\citep{nuscenes2019}, nuScenes is one of the largest open-access datasets available for AV research. The vehicle containing sensors (ego vehicle) in nuScenes' data collections is equipped with 6 cameras, 1 LiDAR\footnote{Light Detection and Ranging sensor} and 5 RADAR, GPS, and IMU\footnote{Inertial Measurement Unit sensors} sensors as the entire sensor suite of an AV. With around 5.5 hours of driving in congested urban areas of Boston and Singapore, nuScenes is a suitable match for studying modern urban spaces. Twenty-three classes of objects are annotated in the dataset, including pedestrians, children, bicycles, construction zones, etc. High-quality annotations of pedestrians make it easier to extract relevant frames from the dataset and focus on the pedestrian side of the scenes. With detailed information on the ego vehicle (the vehicle containing the sensors) position, the dataset enables finding the vehicle-to-pedestrian distance at each time frame. Although the dataset is inclusive of different weather conditions, vegetation, and road markings, such information is not provided in the dataset and needs to be extracted by processing frames and videos. Including the underlying maps of the ego vehicle is another advantage that the nuScenes dataset provides, enabling extracting some context from the map.

Built upon the nuScenes database schema, Lyft level 5 dataset provides 2.5 hours of automated driving in Palo Alto, California~\citep{lyft2019}. Similar to nuScenes, underlying maps, annotations and different weather conditions are included in the dataset. Although it is relatively new, using the same database format as nuScenes gives Lyft level 5 dataset a robust and well-documented structure.  

Google's Waymo dataset is another large and annotated AV dataset publicly available~\citep{sun2020scalability}. With  1,950 scenes of over 6 hours of driving, the covered 76 $km^2$ area in the Waymo dataset is the largest among all the available datasets. Using three coordinate systems and providing means to transform data between frames, it is easy to follow and extract the trajectories of objects by having the positions of objects both in global coordinates and vehicle frame (relative to the ego vehicle position). Extensive hours of data collection include driving in various scenarios, including nighttime and daylight, construction areas, downtown and suburban areas, and diverse weather conditions. The recordings have been captured in Phoenix, Mountain View and San Francisco, enabling research opportunities in domain adaptations. The database format used for Waymo is new and different from those of other datasets, making the application of the models designed based on other datasets require further data cleaning procedures. However, having an active GitHub community with strong documentation helps smooth this transition. Unlike other popular AV datasets, the Waymo dataset currently does not include an underlying map of the events, making utilization of semantic information of the map in the algorithms impossible. 

Finally, PIE (Pedestrian Intention Estimation) dataset was investigated as an AV dataset specifically focused on pedestrians~\citep{rasouli2019pie}. Over 6 hours of annotated video footage of driving in Toronto, Canada, including over 1,800 pedestrian samples with annotated attributes and behaviours, makes PIE a relevant dataset for our research objective. Being limited to camera information, and not including LiDAR data, is the main drawback of the PIE dataset, making it difficult to measure the distance of the objects to the ego vehicle. 
\cref{tab:avcompar} summarizes different features of the reviewed AV datasets.

\begin{table}[!h]
\centering
\caption{{AV dataset comparison}}
\label{tab:avcompar}
\begin{tabular}{lllll|}
\cline{2-5}
\multicolumn{1}{l|}{}                                                                  & \textbf{NuScenes} & \textbf{Lyft}             & \textbf{Waymo} & \textbf{PIE} \\ \hline
\multicolumn{1}{|l}{\textbf{Dimensions:}}\\
\hline
\multicolumn{1}{|l}{Scenes}                                                            & 1000              & 366                       & 1950           & 36             \\ \hline
\multicolumn{1}{|l}{Duration (hr)}                                                     & 5.5              & 2.5                       & 11            & 6           \\ \hline
\multicolumn{1}{|l}{Coverage ($km^2$)}                                                   & 5                 & NA                        & 76             & NA             \\ \hline
\multicolumn{1}{|l}{\textbf{Labelling:}}
\\ 
\hline
\multicolumn{1}{|l}{Annotations}                                                       & 1.4 M             & 1.3 M                       & 12 M           & 3 M         \\ \hline
\multicolumn{1}{|l}{\begin{tabular}[c]{@{}l@{}}Pedestrian \\ Annotations\end{tabular}} & 80 K              & 25 K & 2.8 M          & 800 K           \\ \hline
\multicolumn{1}{|l}{\textbf{Environment:}}\\
\hline
\multicolumn{1}{|l}{Night/Rain/Snow}                                                   & Yes               & No                        & Yes            & No             \\ \hline

\multicolumn{1}{|l}{Maps}                                                              & Yes               & Yes                       & No             & No            \\ \hline
\multicolumn{1}{|l}{\textbf{Sensors:}}\\
\hline
\multicolumn{1}{|l}{LiDAR}                                                              & 1               & 2                       & 5             & No            \\ \hline
\multicolumn{1}{|l}{Camera}                                                              & 6               & 7                       & 5             & 1            \\ \hline
\multicolumn{1}{|l}{RADAR}                                                              & 5               & No                       & No             & No            \\ \hline
\multicolumn{1}{|l}{GPS}                                                              & Yes               & Yes                       & Yes             & Yes            \\ \hline
\multicolumn{1}{|l}{IMU}                                                              & Yes               & No                       & No             & No            \\ \hline
\end{tabular}

\end{table}

To study pedestrian behaviour in unsignalized crossings, we investigate the four datasets mentioned above to extract the relevant frames to the objective of this study.

\subsubsection{Pedestrian crossing in AV datasets}
\begin{figure}
    \centering
    \includegraphics{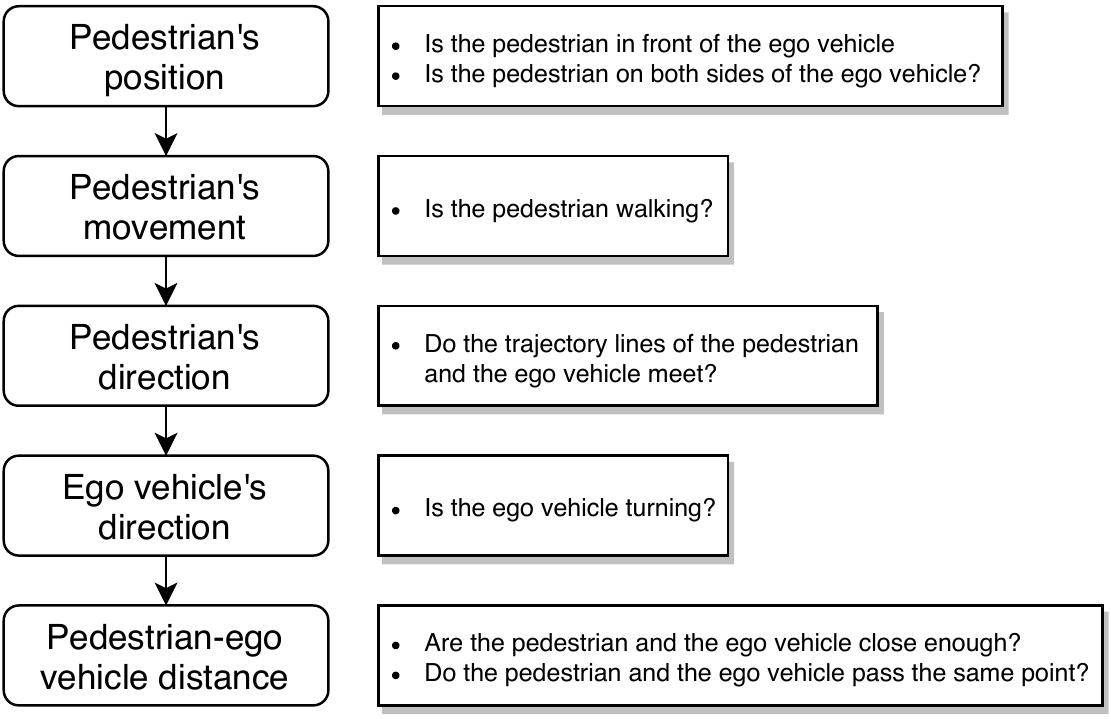}
    \caption{The criteria for extracting mid-block crossings}
    \label{fig:avdiag}
\end{figure}

We extracted pedestrian instances based on the LiDAR data of Lyft, nuScenes and Waymo datasets, and the annotations provided. As the PIE dataset did not include LiDAR data, the pedestrian labels provided in the dataset are used to extract the required information. For the first three datasets, we defined seven criteria and applied them to the datasets to narrow down to events involving mid-block crossings of pedestrians. The defined criteria are:
\begin{enumerate}
    \item Pedestrian is detected in front of the car: as LiDAR data enables detection of pedestrians even if they are on the rear side, a proportion of pedestrians detected are not interacting with the ego vehicle when during the scene. This filter makes sure that the ego vehicle is or will interact with the pedestrian during the scene
    \item Pedestrian is seen on both left and right sides of the ego vehicle: to limit the event to pedestrian crossings, it is expected that during the scene, the pedestrian is detected on both sides of the ego vehicle
    \item Pedestrian is moving in front of the ego vehicle: remove instances of pedestrians waiting, walking on the sidewalk, sitting, etc.
    \item Trajectory of detected pedestrians form an angle of 45 to 135 degrees with the trajectory of the ego vehicle
    \item Ego vehicle's change of direction forms an angle smaller than 60 degrees. This filter is added to avoid turning vehicles to be included as they might pass all the previous criteria with no crossing of a pedestrian taking place
    \item The distance between a pedestrian and the ego vehicle is less than 50 meters
    \item Pedestrian and ego vehicle have intersecting trajectories meaning that they path a similar point during the scene
\end{enumerate}
A diagram of the criteria defined is presented in \cref{fig:avdiag}. It should be noted that due to the short length of scenes, the last criteria should be loosened as the ego vehicle might not necessarily pass the points that pedestrian is observed. The scenes related to remained data are then manually observed to verify if a pedestrian is crossing mid-block in front of the ego vehicle. In the PIE dataset, the pedestrian behavioural labels provided in the dataset include crossing pedestrians, and spatial annotations enable separating pedestrian trajectories based on the type of the cross. Thus, mid-block unsignalized crossings of pedestrians can be extracted using the provided labels in the dataset without any further analysis.

We start with the Lyft dataset, as the smallest dataset of all. In total, 350 scenes were available within the original dataset. Not considering the distance, and applying the first five filters, 137 possible instances of jaywalking pedestrians are found. However, when the instances are limited to a 50-meter maximum distance between pedestrian and ego vehicle, only 20 instances remain. After carefully watching the 20 remained instances, it appeared that no jaywalking events occurred within the dataset. The relatively short 2.5 hours length of the Lyft dataset can be mentioned as the reason behind the failure to find any instances. 

Moving forward to the nuScenes dataset with the same format as Lyft level 5, 5.5 hours of driving is available in the nuScenes dataset, leading to a compressed size of 350 GB. After tracking and detecting pedestrians among different frames of scenes using annotation IDs, 8,143 unique pedestrians were found and tracked in the dataset. After applying all the criteria defined, 200 of the instances remain in the dataset. However, by viewing the videos of the remaining instances, it appears that further filters are required in order to extract jaywalking instances more accurately. For example, in some instances, the ego vehicle is interacting with a pedestrian in a parking lot or a private driveway. Although such instances meet the criteria defined in our filters, they cannot be categorized as mid-block crossing events. Having more information about the environment that the ego vehicle is driving can help distinguish such instances automatically without the need for manual subjective observations or complex video processing. 

The Waymo dataset was the next dataset investigated to extract crossing pedestrian events. In the first step, an impressive number of 23,056 unique pedestrians were tracked between the frames. However, only 1,182 of the total pedestrians passed filter 1 and were detected in front of the car. By applying filter 2,280 of the remaining pedestrians passed the criteria of being observed on both sides of the ego vehicle. One hundred of the remaining pedestrians were removed from the data by adding the walking filter, leading to 211 instances of potential mid-block crossings. By applying the angle criteria, 117 pedestrians were left in the dataset. However, when observing the video data of the remaining instances, it appeared that a major part of the crosses was related to ego vehicle turning events, which made the walking pedestrians on the sidewalks pass all the previous filters. By introducing filter 5 and focusing on vehicles following a relatively straight trajectory, only 55 instances were left in the dataset.

Finally, PIE dataset, which concentrates on pedestrians and provides the labels required to study them was explored. In total, PIE includes 1,842 instances of pedestrians, 517 of which are related to crossing pedestrians. After cleaning the data and limiting the instances to unsignalized mid-block crossings, 47 instances of cross remained in the dataset.

By investigating some of the most well-established open-access AV datasets, it can be concluded that in order to extract events of a specific behaviour of pedestrians, larger and more extensive datasets are required. A solution to overcome this challenge would be to combine data from different resources to create a hybrid dataset focused on pedestrians. However, the differences in data formats, sensors, environments and coordinate systems used make it a challenging and difficult task to combine these datasets. Another solution would be to collect data with a particular focus on pedestrians. However, most popular video datasets used for pedestrian behaviour analysis are dedicated to pedestrian interactions with each other and their dynamics in crowds~\citep{robicquet2016learning,zhang2019widerperson}. In recent years, researchers have tried to address this issue and collect and provide pedestrian-oriented AV datasets~\citep{kotseruba2016joint}. However, these datasets are still not well-established in the literature. To understand essential context information required for pedestrian trajectory studies, we developed a controlled VR experiment. The controlled nature of our data allows us to record pedestrian behaviour under several customized conditions and test the effects of various contextual information on model accuracy.    

\section{Virtual Reality Data}
\label{S:t3}
Lack of inclusiveness for pedestrians and, in particular, crossing pedestrians in the available AV datasets raises the need for exploring sources specifically designed to account for behavioural patterns of jaywalking and mid-block crossing pedestrians. One solution to achieve such data in a controlled environment, with the possibility of customizing scenarios to include desired conditions, would be to develop and conduct Virtual Reality (VR) experiments. By doing so, and analyzing the behaviour in a controlled, safe, and relatively low-cost environment, we can acquire information on factors determining pedestrian trajectory and provide solutions and suggestions for improving and complementing AV datasets.

For this study, Virtual Immersive Reality Environment (VIRE) is used to simulate a range of different scenarios and conduct experiments on different aspects of pedestrian crossing and walking behaviour. Introduced by~\cite{farooqvire}, VIRE uses Head Mounted Display and virtual reality to enable interactive, immersive and complex simulated scenarios. Hypothetical traffic scenarios can be projected directly to the eyes of users, and with a human-in-the-loop simulation, the behaviour of simulated vehicles is influenced by the participants' behaviour. In this study, scenarios involve a mid-block crosswalk, with vehicles passing by on the street. Each scenario is designed by a set of variables with different levels designed. Participants wearing the VR headset start on the simulated sidewalk, and in different conditions, are asked to cross the street when they feel it is safe to do so. While performing the experiment, pedestrian reactions, including their coordinates, head orientation and distance to the approaching vehicles, are recorded in 0.1-second intervals.   

\begin{figure}[!h]
\centering
\begin{subfigure}{0.8\linewidth}

    \begin{subfigure}[b]{0.5\textwidth}
         \centering
         \includegraphics[width=0.91\textwidth,height=3.5cm]{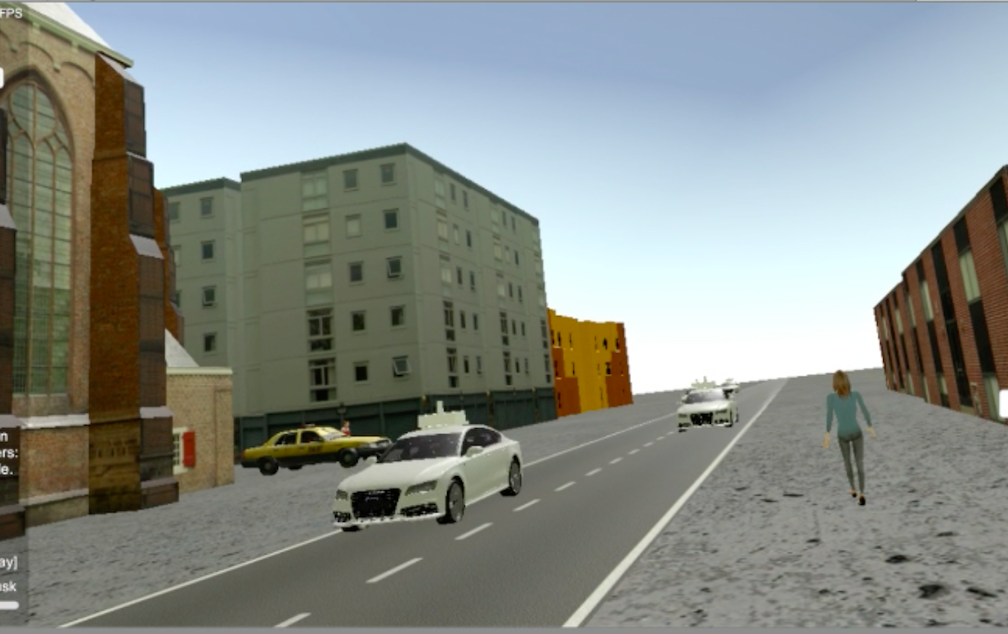}
         \caption{A sample day view of the environment}
     \end{subfigure}
     \hfill
     \begin{subfigure}[b]{0.5\textwidth}
         \centering
         \includegraphics[width=0.91\textwidth,height=3.5cm]{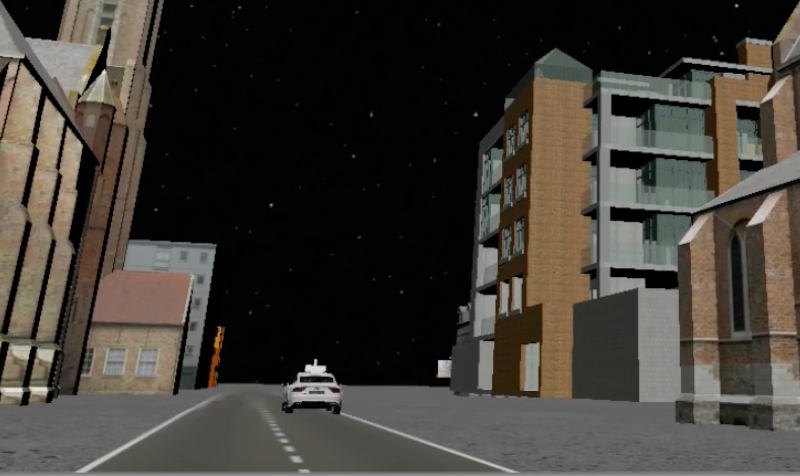}
         \caption{A sample night view of the environment}
     \end{subfigure}
\end{subfigure}\\[2ex]
\begin{subfigure}{0.8\linewidth}

\centering
\includegraphics[scale=0.107]{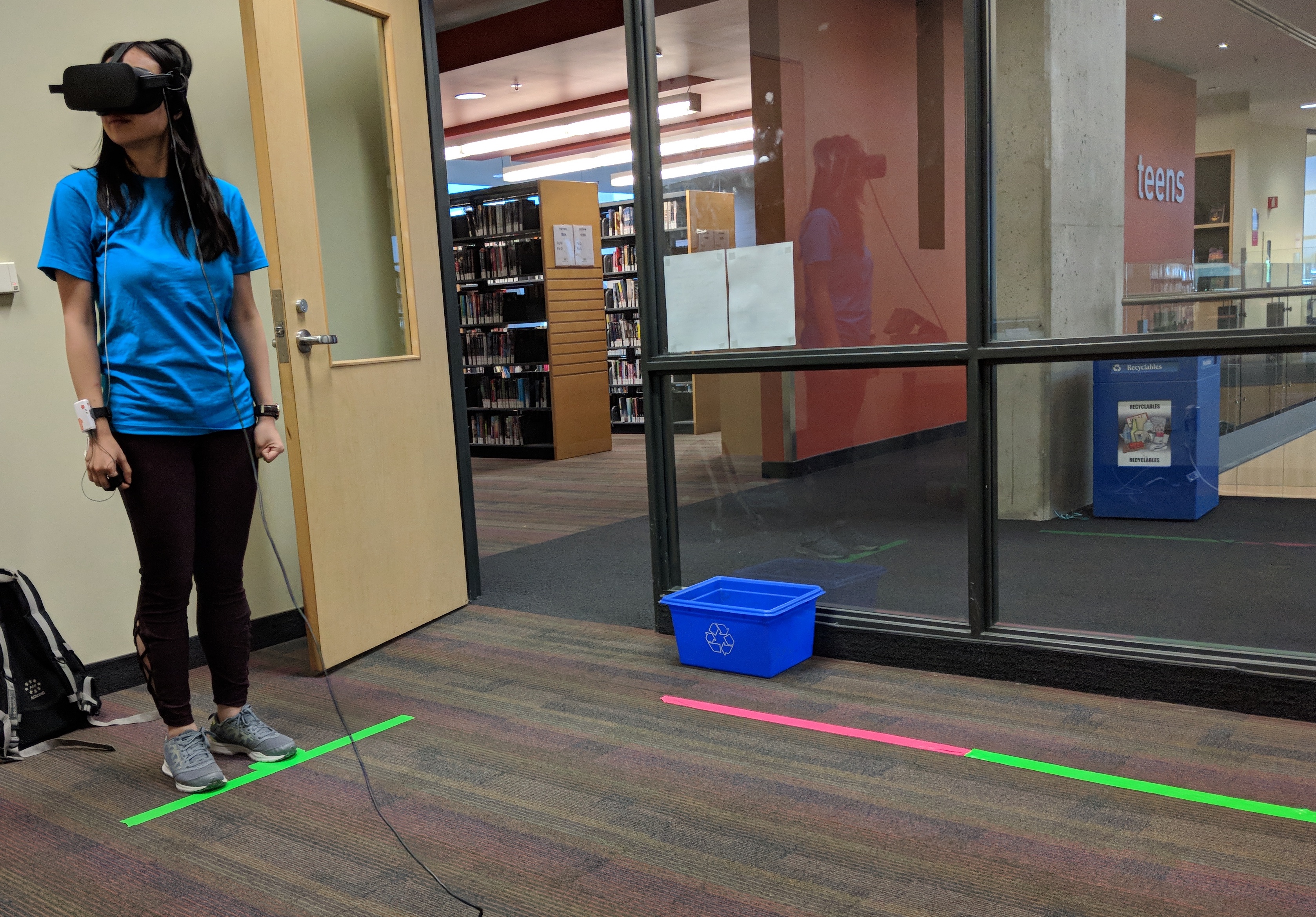}
\caption{A participant doing the experiment}
\end{subfigure}

\caption{Images from a virtual reality experiment}
\label{fig:Texp}
\end{figure}

The data collection campaign was conducted in the summer of 2018 in four different locations to cover a heterogeneous population. A total of 180 individuals from different age groups participated in the experiment over a period of 5 months. The experiments were performed at Ryerson University, City of Markham Public Library, Toronto City Hall, and North York Civic center. Participants were exposed to multiple scenarios, with changing parameters in each round. In \cref{fig:Texp}, an experiment and two sample views of the VR environment are shown.

In addition to the recorded time-series data from participants, contextual variables from the scenario's environment were captured to include in trajectory prediction models. The context variables include the type of road (one-way, two-way or two-way with median), speed limit (30, 40 or 50 km/hr), lane width (2.5, 2.75 or 3 m), weather conditions (snowy day or clear view), time of the day (day or night), and arrival rate of cars (530, 750, 1100 veh/hr). These variables are selected so that a hypothetical AV can capture and utilize them as input to its trajectory prediction algorithm. Detailed information on the data collection campaign, collected data, scenario details, etc. can be found in ~\cite{kalatian2020decoding}. To concentrate on the interactions of pedestrians with AVs, scenarios involving simulated human-driven vehicles in the traffic are not included in this study. Also, other scenario variables that cannot be captured by cameras or LiDAR sensors are not used in the modelling to ensure that a hypothetical AV can deploy the proposed algorithms without requiring information that they cannot capture inherently.
\section{Methodology}
\label{S:t4}
Pedestrian movement patterns are highly correlated both temporally and spatially~\citep{song2016deeptransport}. Recurrent Neural Networks (RNNs) are a popular choice to deal with the problem by treating mobility patterns of a pedestrian as a sequence prediction problem. However, it has been shown that RNNs are not capable of remembering long-term temporal and spatial dependencies as a result of the problem of vanishing gradient~\citep{hochreiter1997long}. Introduced in \citep{hochreiter1997long}, Long Short-Term Memory (LSTM) is a modification to traditional RNN architecture that enables learning sequence labels for longer time intervals by implementing four interactive gates. 


In this study, we propose, \textit{Aux-LSTM}, a novel framework consisting of multi-input LSTM layers and fully connected dense layers, to predict the next coordinates of pedestrians as output. \arash{Initial steps of time-series data, i.e., coordinates ($x_0,y_0$), head orientations ($o_0$), and distance to vehicle ($d_0$), are used as input to the LSTM layers. The output of the LSTM layers will then merge with extra information from contextual variables ($C$), and the mergers enter a series of fully connected dense layers to predict the pedestrians' future trajectory ($x_f,y_f$) in the rest of their crossing. In the mathematical form, the neural network architecture to predict the future trajectory of a pedestrian ($T = \{x_f,y_f\}$)  using time-series data ($S = \{x_0, y_0, o_0,d_0\}$) is as follows:}
\begin{itemize}
    \item LSTM hidden layer(s): 
    \begin{equation}
        H_l = L(S, W_l,\sigma)
    \end{equation}
    \item Dense hidden layer(s) variables
    \begin{equation}
    H_d=f(H_l,C, W_d,\sigma)
    \end{equation}
    \item Output layer:
    \begin{equation}
    T = f(H_d, W_o,\phi)
    \end{equation}
\end{itemize}
\arash{where $W_l,W_d,W_o$ are the weights of hidden LSTM, hidden dense and output dense layers, $\sigma$ and $\phi$ are activation functions with ReLU and Sigmoid nonlinearities, and $L$ and $f$ are LSTM and Dense layers, respectively.}

\begin{figure}[h]
    \centering
    \includegraphics[scale=0.7]{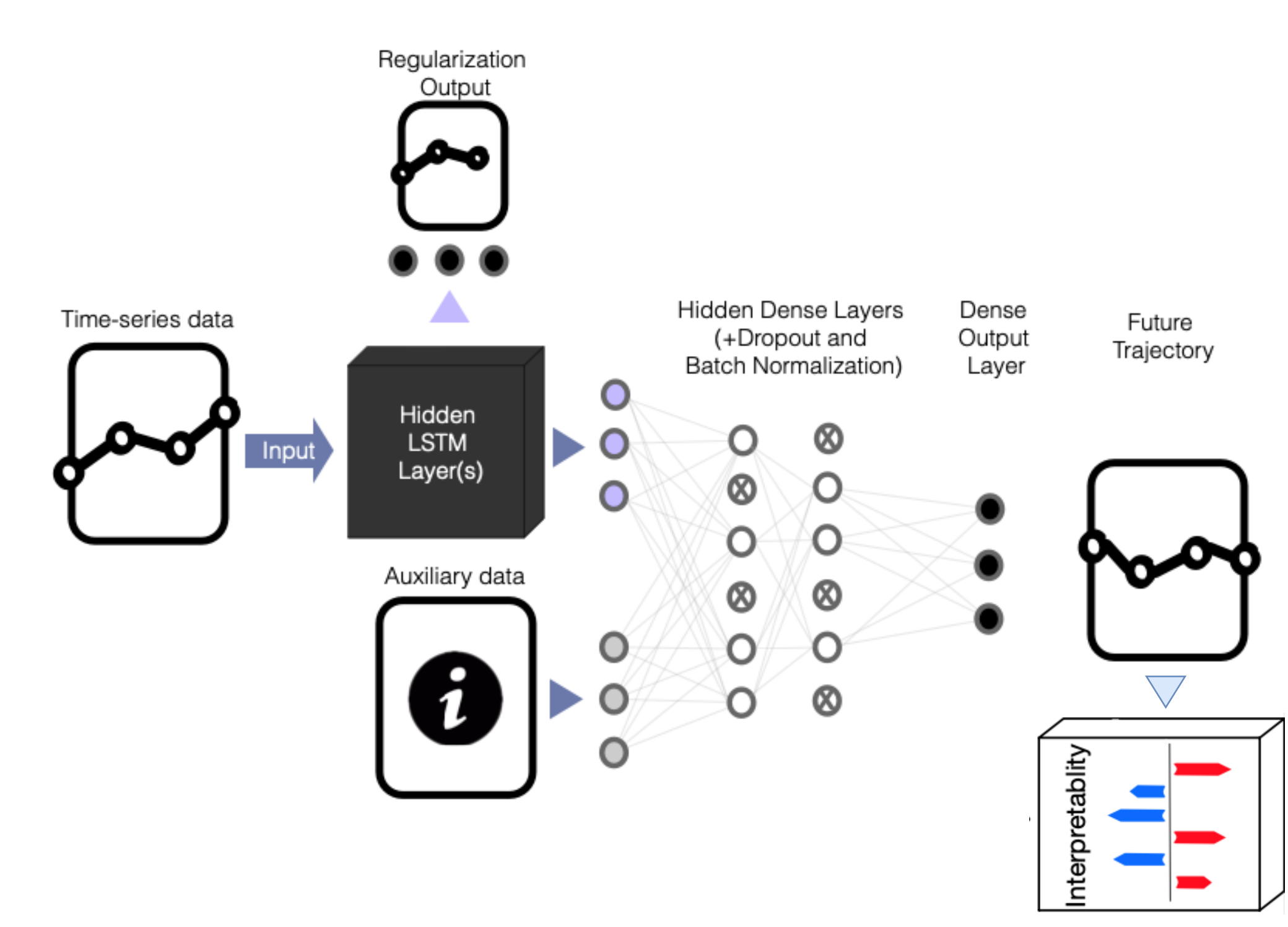}
    \caption{Schematic framework of Aux-LSTM}
    \label{fig:Tframe}
\end{figure}

As a regularization mechanism to the framework, the model is supervised through two identical loss functions. Both loss functions are defined as the mean squared error of the predicted coordinates. By using the loss function after the LSTM layers, a.k.a. secondary loss, we allow smoother training for the framework. Batch Normalization and Dropout layers are also used in order to reduce overfitting in the model. \cref{fig:Tframe} depicts the general framework of Aux-LSTM.

Input time-series data are defined in two ways: \textit{time-based} and \textit{distance-based}. In the time-based approach, the pedestrian's coordinates in the next $t_2$ seconds are predicted based on their last $t_1$ seconds of behaviours. At each point during the cross, pedestrian coordinates, head orientations, and their distance to the approaching vehicle during the last $t_1$ seconds are used as time-series input to predict the coordinates of the pedestrian in the next $t_2$ seconds. In the distance-based type of models, however, the proportion of data used as input, $p$, is defined as the proportion of lane width that the pedestrian has passed when the algorithm tries to predict the pedestrian coordinates in the rest of the cross. For instance, if $p$ is set to 0.3, the framework tries to predict the pedestrians' trajectory based on their trajectory in the first 30\% of the lane width. Different values of $t_1$, $t_2$ and $p$ are tested in order to provide insights into the required method of data preparation. In the AV context, a time-based approach enables the continuous observance of pedestrian behaviour and prediction of their next movements. A distance-based approach, on the other hand, makes it possible to predict the whole trajectory of the pedestrians in front of the vehicle to the point when they fully cross the road. It should be noted that time-based models and distance-based models are not meant to be competing as each serves to provide answers to a different question. In time-based models, it is assumed that the vehicle observes the movements of the pedestrians continuously, and thus it can periodically update the coordinates with ground truth information. On the other hand, distance-based models aim to provide insights into the performance of the model on longer sequences. For practical purposes, time-based models are used and compared in the relevant literature.

\arash{One of the main barriers to even more prevalent use of neural networks, especially in practical applications, is the difficulties involved in their interpretability and their \textit{blackbox} nature. As this study suggests Virtual Reality data as a tool to complement AV datasets, providing insights on the contributing factors to the error in trajectory prediction of pedestrians can be beneficial to future AV data collections, as well as current AV data analysis. In this study, SHAP~\citep{lundberg2017unified}, a post-hoc game theory-based interpretation method is utilized to understand the effect of variables on the prediction error.}

\arash{Shapley value is a method rooted in game theory to allocate the payoff of a job done in a coalition, to the involved contributors. The same concepts can be applied in determining the contribution of each variable to the model output. Shapley value of each variable $i$ is calculated as follows~\citep{lipovetsky2001analysis}:}

\begin{linenomath}
\begin{equation}
\label{eq:shapley}
\Phi_i = \sum_{S \in F \setminus \{i\}} \frac{|S|! (|F| - |S| -1)!}{|F|!} \left(
g_{S\cup\{i\}}(x_{S\cup\{i\}}) - g_S(x_S) \right)
\end{equation}
\end{linenomath}
\arash{In \cref{eq:shapley}, $S$ is a subset of all features ($F$), $g_{S\cup\{i\}}$ is the model trained using a subset with feature $\{i\}$ present, and $g_S$ is the model trained without the feature $i$. Similarly, $x_{S\cup\{i\}}$ and $x_{S}$ represent the values of input features in subset $S$ when feature $i$ is and is not present, respectively.  Lundberg~\textit{et al.}~\citep{lundberg2017unified} applied the concept of Shapley values to model interpretation and proposed to estimate the importance of a variable in an instance based on the corresponding Shapley values.}

\section{Results and Analysis}
\label{S:t5}
In this section, we discuss the results of applying the proposed framework to the virtual reality dataset. The two input data formats are presented and analyzed to test the performance of prediction in different conditions. Moreover, in order to test the applicability of the models to real-world video data, the model is also trained on pedestrian trajectories at mid-block crosses extracted from PIE dataset. Finally, a game theory-based machine learning interpretability method is applied to the model trained on the VR data to assess the contribution of contextual variables to the model accuracy.  
\subsection{Implementation Details}
All data pre-processing and model development are coded in Python programming language, using Keras library and its implementation of TensorFlow with GPU support. After the data preparation process, an exhaustive grid search is conducted to find the best network configurations. Dropout layers and their rates, number of nodes (neurons) in each hidden layer, batch size, number of hidden LSTM and dense layers are configured based on 8-fold cross-validation over 100 epochs, and the best configurations are selected to test on a separate test dataset. Models are trained on a Core i7 4 GHz CPU and a 16.0 GB memory.

In total, 3,276 instances of cross were collected using the virtual reality experiments, which is significantly higher than the mid-block crossing events detected using all the open-access AV datasets currently available. To compare the order of magnitude, the most relevant open-dataset, PIE, includes 47 such instances. For distance-based models, each instance of cross is divided into two parts: 1. input to feed the time-series part of the networks and 2. output to be predicted. Three input proportions ($p$) are tested as the proportion of the length of the road that shapes the input data: 0.3, 0.5, 0.7. \arash{A larger $p$ means that the vehicle predicts the next movements of the pedestrians based on longer observations of their behaviour. Therefore, the models with larger $p$ as input are expected to perform more accurately. On the other hand, in cases of higher speeds, vehicles might not have enough time to react timely in a larger $p$.} The corresponding distance-based data types are $D\_3,$, $D\_5$ and $D\_7$, respectively. In time-based models, on the other hand, each instance of crossing is converted to multiple sequences. Based on model parameters, each $t_1$ second of the pedestrian trajectory is used as input to the time-series part of the data, and the following $t_2$ seconds are used as the output to be predicted using the model. Hence, in time-based models, data size increases significantly compared to distance-based models. Three combinations of sequence duration lengths are tested in this study. In the first generated data of this type, $T\_1\_1$, each 1 second of the pedestrian's recorded behaviours are used to predict the next 1 second. $T\_1\_2$ and $T\_2\_1$ are the two other time-based data types used to train time-based models in this study, with a $t_1$ of 1 and 2 seconds and a $t_2$ of 2 and 1 seconds, respectively. By using various proportions and formats of the input data, we tried to understand the performance of our proposed algorithm under different scenarios. \cref{tab:datasize} provides the number of samples used for training and testing each of the models, generated from the 3,276 crossing instances. All the models are trained and validated using 80\% of the data and tested on the remaining 20\%.

A grid search of hyperparameters is conducted to find the best-performing model of each type. Parameters investigated in the search include batch sizes (32, 64 and 128), dropout rates (0, 0.2 and 0.5), number of nodes (10, 50 and 100), number of hidden LSTM layers(1, 2, and 3) and number of hidden dense layers (1, 2 and 3).      
\begin{table}[!h]
\centering
\footnotesize
\caption{Number of samples of VR data used for training and testing for each data type}
\begin{tabular}{|ll||ll|}
\hline
\textbf{Distance-based} & \textbf{Samples} & \textbf{Time-based} & \textbf{Samples} \\ \hline
$D\_3$                     & 3,261             & $T\_1\_1$               & 58,654            \\ \hline
$D\_5$                     & 3,261             & $T\_1\_2$               & 32,455            \\ \hline
$D\_7$                     & 3,261             & $T\_2\_1$               & 32,455            \\ \hline
\end{tabular}
\label{tab:datasize}
\end{table}

\subsection{Baseline Model}
Vanilla LSTM models are used as the baseline model to assess the performance of Aux-LSTM models. The data processing procedure and other configuration setup steps for the baseline models are similar to the Aux-LSTM models. 
To find the best combination of the time-series information to be fed into the models, four variants of the models are trained and compared. Time-series input include: Participants' coordinates $\{x_0, y_0\}$, head orientations $\{o_0\}$ and distance to vehicles $\{d_0\}$. To find the best combination, four variants of the models are defined and trained as follows:
\begin{itemize}
    \item \textbf{Variant-xy}: receives solely pedestrians' coordinates ($x_0,y_0$) as time-series input
    \item \textbf{Variant-xyo}: receives pedestrians' coordinates ($x_0,y_0$) and head orientations ($o_0$) as time-series input
    \item \textbf{Variant-xyd}: receives pedestrians' coordinates ($x_0,y_0$) and distance to vehicles ($d_0$) as time-series input 
    \item \textbf{Variant-xyod} receives pedestrians' coordinates ($x_0,y_0$), head orientations ($o_0$) and distance to vehicles ($d_0$) as time-series input.
\end{itemize}
For each variant, distance-based and time-based models as discussed in the next subsections are developed, and the performance of the models is compared based on the error on the validation set to find the best configuration. Test sets are used in the final step for the final model performance assessment. The loss function to be minimized during the training of the models is defined as the Mean Square Error (MSE) of the predicted and ground truth values for the coordinates followed by the pedestrian. Moreover, the Root Mean Square Error (RMSE) of the difference between predicted and ground truth coordinates is used as the indicator of the performance of the models. Based on the results of the top-performing models of the four variants, it appears that the addition of head orientation and distance to vehicle to the coordinates of pedestrians improves the accuracy in predicting the pedestrians' future trajectory. As this information can be obtained with relatively low costs in real-world AVs by either their sensors or cameras, they can be collected periodically and used to enhance the accuracy of trajectory prediction models. In the next sections, Variant-xyod is further investigated. The cross-validation results of all four variants are provided in \ref{A:valT}.  
\subsection{Distance-based Models}
\cref{tab:distmodelsxyodm} presents the configurations of top distance-based Vanilla and Aux-LSTM models based on 3 different values of $p$, the proportion of lengths of the road that its corresponding time-series data are used as input to the model. As shown in this table, adding dropout layers has not contributed to the better-performing models, and in all of the 6 top configurations of different models, the dropout rate has appeared to be 0. Also, a general trend of a decrease in the depth and density of the networks by the increase in the length of the input sequence can be observed.  
\begin{table}[!h]
\centering
\caption{8-fold cross-validation results for top distance-based models trained on VR data: Variant-xyod. Errors are reported in meters as root mean square error over all predicted time steps}
\label{tab:distmodelsxyodm}
\scalebox{0.7}{
\begin{tabular}{|lllllllllll|}
\hline
\textbf{p}           & \textbf{Type} & \textbf{\begin{tabular}[c]{@{}l@{}}Dense\\ Layers\end{tabular}} & \textbf{\begin{tabular}[c]{@{}l@{}}LSTM \\ Layers\end{tabular}} & \textbf{Nodes} & \textbf{\begin{tabular}[c]{@{}l@{}}Batch \\ Size\end{tabular}} & \textbf{Dropout} & \textbf{\begin{tabular}[c]{@{}l@{}}Validation \\ Loss \end{tabular}} & \multicolumn{1}{l}{\textbf{\begin{tabular}[c]{@{}l@{}}Validation \\ Error \end{tabular}}} & \textbf{\begin{tabular}[c]{@{}l@{}}Train \\ Loss \end{tabular}} & \textbf{\begin{tabular}[c]{@{}l@{}}Train \\ Error \end{tabular}} \\ \hline
\multirow{2}{*}{0.3} & Aux-LSTM      & 2                                                               & 3                                                               & 100            & 32                                                             & 0                & 0.0219                                                                    & 0.1481                                                                                        & 0.0171                                                               & 0.1306                                                              \\
                     & Vanilla       & NA                                                              & 2                                                               & 100            & 32                                                             & 0                & 0.0180                                                                    & 0.1341                                                                                        & 0.0183                                                               & 0.1351                                                              \\ \hline
\multirow{2}{*}{0.5} & Aux-LSTM      & 2                                                               & 2                                                               & 50             & 128                                                            & 0                & 0.0104                                                                    & 0.1021                                                                                        & 0.0115                                                               & 0.1071                                                              \\
                     & Vanilla       & NA                                                              & 1                                                               & 50             & 128                                                            & 0                & 0.0088                                                                    & 0.0940                                                                                        & 0.0071                                                               & 0.0841                                                              \\ \hline
\multirow{2}{*}{0.7} & Aux-LSTM      & 1                                                               & 3                                                               & 50             & 32                                                             & 0                & 0.0069                                                                    & 0.0830                                                                                        & 0.0059                                                               & 0.0769                                                              \\
                     & Vanilla       & NA                                                              & 1                                                               & 10             & 32                                                             & 0                & 0.0025                                                                    & 0.0500                                                                                        & 0.0030                                                               & 0.0545                                                              \\ \hline
\end{tabular}}
\end{table}

In addition to model configurations, loss and error over validation and training data are provided in \cref{tab:distmodelsxyodm}. Comparing Aux-LSTM models with the baseline models, it can be observed that validation errors of all the vanilla models are less than their Aux-LSTM counterparts. The difference in error ranges from around 0.01 meters in D\_3 model with p = 0.3, to around 0.03 meters in D\_7 with an input proportion of 0.7. It can be observed that in general and based on the validation error, adding auxiliary information has not helped the model perform better. In addition, with an increase in the length of the time-series input sequence, the contribution of auxiliary information is decreased. To assess the performance of the networks more accurately, all the selected trained models are applied to the test dataset. The errors of applying the models on test set over 100 epochs are provided in \cref{tab:Dtest}. According to the table, the performance of all the Aux-LSTM models, when applied to the test set, is better than the vanilla baseline models. Interestingly, the input proportion with the most accurate models over the validation dataset, i.e. p = 0.7, has the most significant gap between its Aux-LSTM and baseline model. The Difference in the performance of the models over validation and test sets can be traced back to the effect of adding auxiliary data to the input sequences and their contribution to the diversification of the input information and to the reduction of relying solely on time-series input data. A bigger difference in errors of models with p = 0.7 confirms the idea that having longer sequences of time-series data as input leads to more reliance of the model on the available time-series input, which is reduced by adding auxiliary information to the input data. In distance-based models with p = 0.3, 0.5 and 0.7, considering auxiliary data in the models has reduced the root mean square error of coordinate predictions by 2\%, 6\% and 17\%, respectively. As stated earlier, the differences between accuracy improvements of the models confirm that relatively longer input sequence lengths benefit to a greater extent when auxiliary data is added.

\begin{table}

    \caption{Mean error of test set in meters for selected distance-based models over 100 epochs trained on VR data} 
    \centering
    \footnotesize
    \addtolength{\tabcolsep}{-3pt}
    \begin{tabular}{|l|l|c|}
    \hline
       \textbf{ID} & \textbf{Model Type}  & \textbf{Error} \\
       \hline
    
         \textbf{P: 0.3} & Aux-LSTM & 0.3284   \\
           & Vanilla & 0.3307\\
          \hline
         \textbf{P: 0.5}  &  Aux-LSTM & 0.2826  \\
           & Vanilla & 0.3019 \\
         \hline
         \textbf{P: 0.7} & Aux-LSTM & 0.2329  \\
           & Vanilla & 0.2795 \\

    \hline
    \end{tabular}
    \label{tab:Dtest}
\end{table}

\subsection{Time-based Models}
Similar steps are followed for time-based models. \cref{tab:timemodels} presents the configurations of top-performing time-based models over validation data, as well as their corresponding validation and training error. It is interesting to note that except for the number of hidden LSTM layers in T\_2\_1, all the other configurations of the top models appear to be similar to each other.

\begin{figure}[!h]%
 \centering
 \subfloat[A distance-based model sample, p : 0.3: 1]{\includegraphics[scale=0.35]{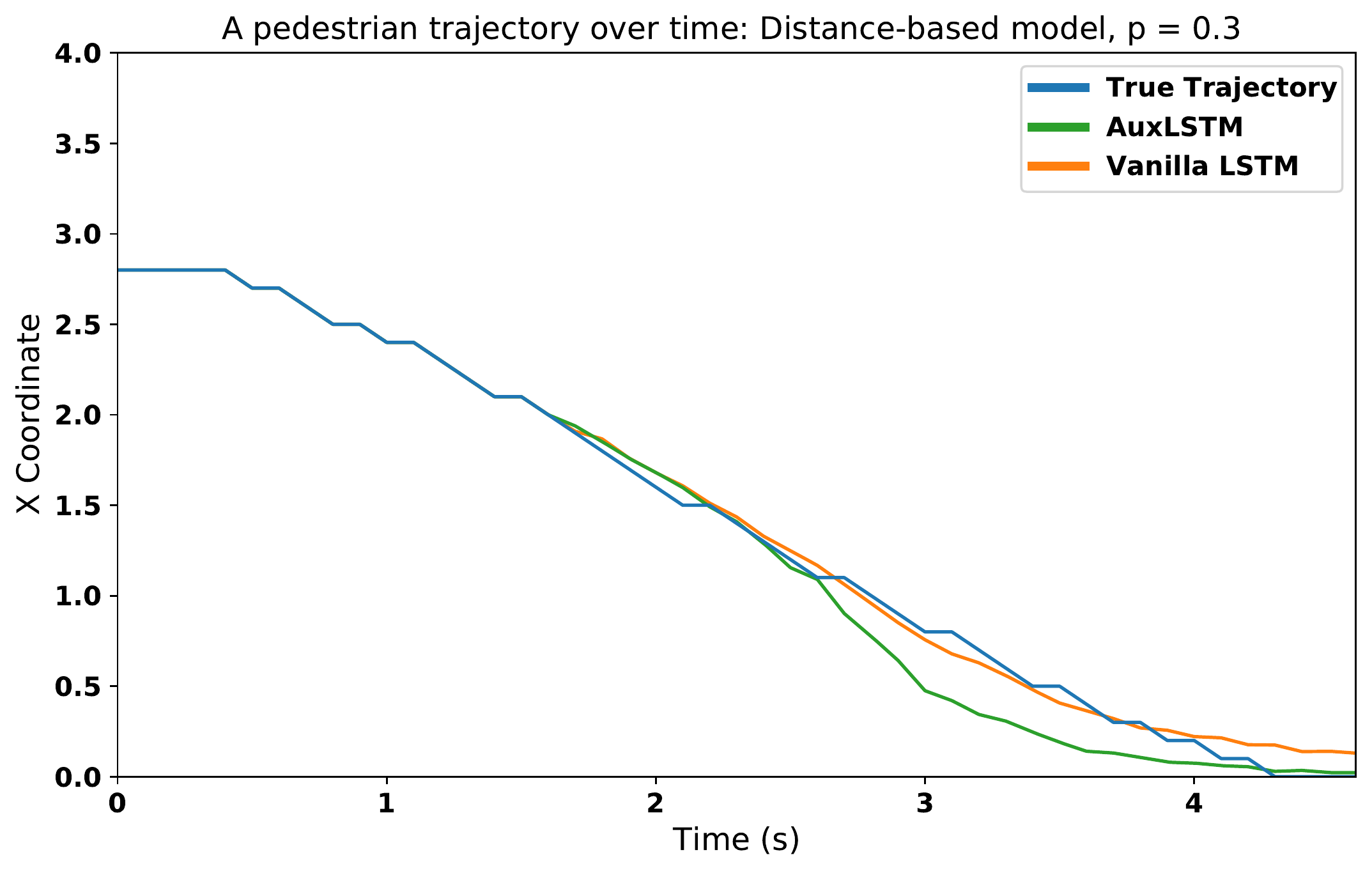}\label{fig:a}} \quad
 \subfloat[A time-based model sample, $t_1$: 1, $t_2$: 1]{\includegraphics[scale=0.35]{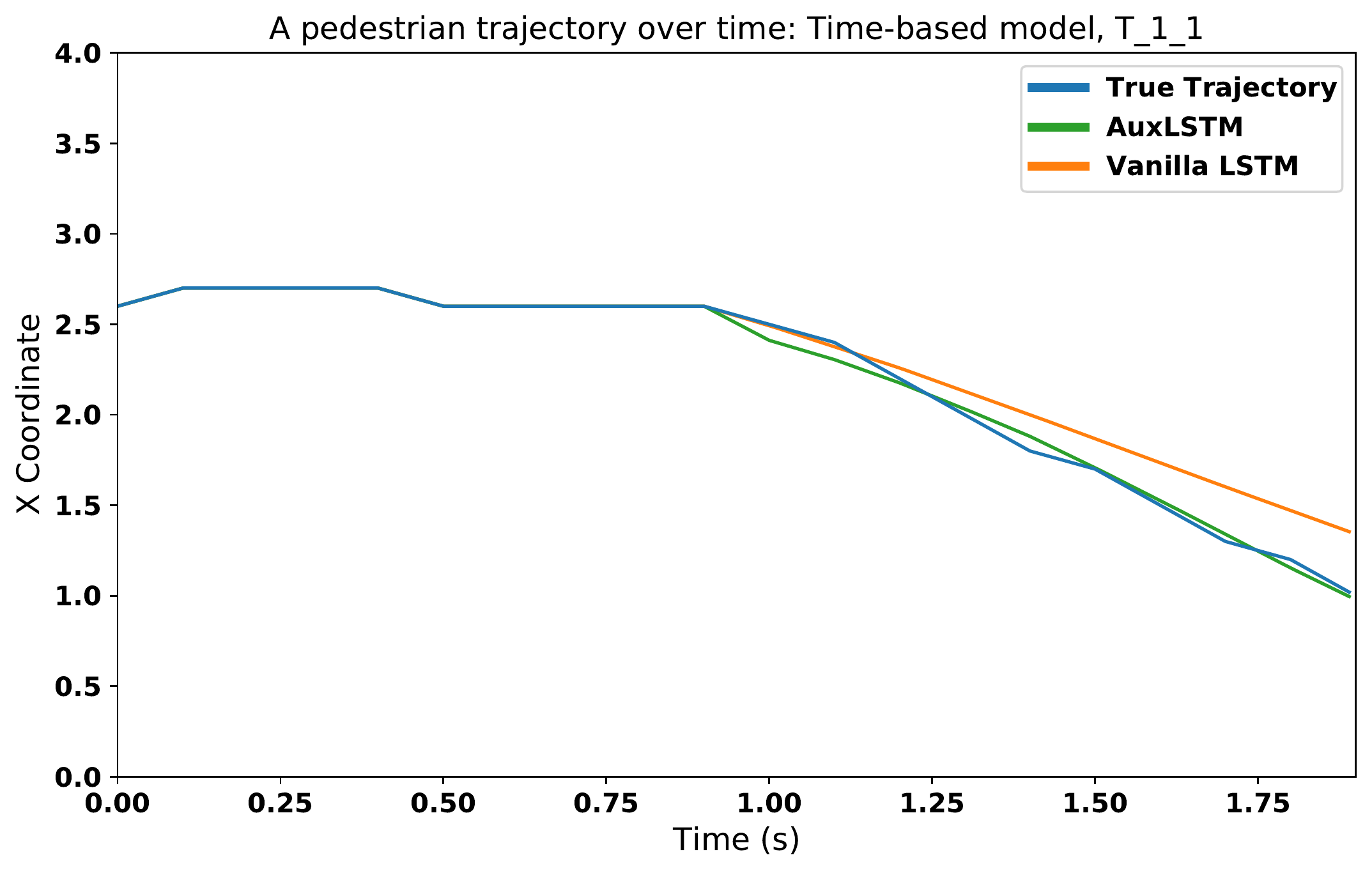}\label{fig:b}}\\
 \caption{Sample user's trajectory and prediction using vanilla and aux-LSTM models}%
 \label{fig:compar}%
\end{figure}

Unlike distance-based models, all Aux-LSTM time-based models outperform the baseline models over the validation data. The same pattern exists when the models are applied to the test set (\cref{tab:Ttest}), 
and the performances of the time-based Aux-LSTM models are still better than the baseline models, which is achieved through a shorter length of sequences as input data. In general, root mean square errors of coordinate prediction using time-based Aux-LSTM models outperform the baseline models over T\_1\_1, T\_1\_2 and T\_2\_1 data types by 7\%, 12\% and 12\%, respectively. Compared to distance-based models, the gap between accuracy improvements of different time-based data types is smaller, which can be traced to the smaller differences in input and output sequence lengths in time-based data. {\cref{fig:compar} depicts two prediction samples from distance-based and time-based models. Ground truth trajectories of sample users from the test set, along with their corresponding predicted trajectories using vanilla and Aux-LSTM models are provided in the figures. The Y axis in the figures corresponds to the X coordinate of the trajectories, which is also equivalent to the distance from a reference point on the other side of the road. In \ref{A:figures}, more sample trajectories from all data types and for different pedestrian speeds are provided (animated versions of samples from all data types are also included in the supplementary materials).}

\begin{table}[!h]
\centering
\caption{8-fold cross-validation results for top time-based models trained on VR data: Variant-xyod. Errors are reported in meters as root mean square error over all predicted time steps}
\label{tab:timemodels}
\scalebox{0.68}{
\begin{tabular}{|llllllllllll|}
\hline
\textbf{$t_1$ }    & \textbf{$t_2$ }    & \textbf{Type} & \textbf{\begin{tabular}[c]{@{}l@{}}Dense\\ Layers\end{tabular}} & \textbf{\begin{tabular}[c]{@{}l@{}}LSTM \\ Layers\end{tabular}} & \textbf{Nodes} & \textbf{\begin{tabular}[c]{@{}l@{}}Batch \\ Size\end{tabular}} & \textbf{Dropout} & \textbf{\begin{tabular}[c]{@{}l@{}}Validation \\ Loss \end{tabular}} & \multicolumn{1}{l}{\textbf{\begin{tabular}[c]{@{}l@{}}Validation \\ Error \end{tabular}}} & \textbf{\begin{tabular}[c]{@{}l@{}}Train \\ Loss \end{tabular}} & \textbf{\begin{tabular}[c]{@{}l@{}}Train \\ Error \end{tabular}} \\ \hline
\multirow{2}{*}{1} & \multirow{2}{*}{1} & Aux-LSTM      & 3                                                               & 2                                                               & 100            & 32                                                             & 0                & 0.0181                                                                    & 0.1344                                                                                        & 0.0085                                                               & 0.0922                                                              \\
                   &                    & Vanilla       & NA                                                              & 2                                                               & 100            & 32                                                             & 0                & 0.0312                                                                    & 0.1767                                                                                        & 0.0170                                                               & 0.1304                                                              \\ \hline
\multirow{2}{*}{1} & \multirow{2}{*}{2} & Aux-LSTM      & 3                                                               & 2                                                               & 100            & 32                                                             & 0                & 0.0305                                                                    & 0.1748                                                                                        & 0.0139                                                               & 0.1178                                                              \\
                   &                    & Vanilla       & NA                                                              & 3                                                               & 100            & 32                                                             & 0                & 0.0781                                                                    & 0.2795                                                                                        & 0.0322                                                               & 0.1795                                                              \\ \hline
\multirow{2}{*}{2} & \multirow{2}{*}{1} & Aux-LSTM      & 3                                                               & 3                                                               & 100            & 32                                                             & 0                & 0.0068                                                                    & 0.0827                                                                                        & 0.0049                                                               & 0.0700                                                              \\
                   &                    & Vanilla       & NA                                                              & 2                                                               & 100            & 32                                                             & 0                & 0.0097                                                                    & 0.0983                                                                                        & 0.0053                                                               & 0.0726                                                              \\ \hline
\end{tabular}}
\end{table}
\begin{table}[!h]
\footnotesize
    \caption{Mean error of test set in meters for selected time-based models over 100 epochs trained on VR data}
    \centering
    \addtolength{\tabcolsep}{-3pt}
    \begin{tabular}{|l|l|l|}
    \hline
       \textbf{ID} & \textbf{Model Type}  & \textbf{Test Error } \\
       \hline
    
         \textbf{$t_1$: 1, $t_2$: 1}  & Aux-LSTM &  0.2606   \\
         \textbf{}  & Vanilla & 0.2801\\
         \hline
         \textbf{$t_1$: 1, $t_2$: 2} & Aux-LSTM & 0.3642     \\
          & Vanilla & 0.4122 \\
         \hline
         \textbf{$t_1$: 2, $t_2$: 1}  & Aux-LSTM & 0.2707\\
            & Vanilla &  0.3083 \\
    \hline
    \end{tabular}
    \label{tab:Ttest}
\end{table}

\subsection{Application to open-access AV dataset}
A major criticism of virtual reality data is the effect of a controlled and safe environment on the behaviour of participants~\citep{kinateder2014virtual}. In order to test the Aux-LSTM framework on a real dataset, \arash{and after a careful investigation of several available datasets,} we selected PIE dataset, as an open-dataset featuring detailed labels of pedestrians. \arash{The pedestrian-oriented data collection in this dataset makes it a great choice for our study, as other similar datasets did not contain adequate instances of pedestrian crossings.} To account for the difference in the provided information in PIE dataset, auxiliary variables used for training the model are re-selected based on the annotations of the dataset. Moreover, PIE dataset does not include LiDAR data, which prevents measuring the distance of the ego-vehicle to the pedestrians. Thus, the coordinates of pedestrians provided in the dataset are based on their relative location in the camera frame. Type of road (one way or two way), number of lanes and the ego vehicle speed when the pedestrian is first detected are used as auxiliary variables. Head orientations and distance to vehicle, which were used as time-series data for the model trained on VR data, are not available through PIE dataset. Speed of the vehicle at each time interval is instead used as input to the LSTM layers, along with coordinates of the crossing pedestrian in the camera frame. It should be noted that PIE dataset includes other behavioural annotations such as gender and age category of pedestrians. However, such annotations are not used in order to avoid human-labelled information, \arash{to make sure that the model can be deployed independently by a typical AV without a need for other sociodemographic predictions, which might add to the error and bias in the model.} . As the distance followed by the objects is not measurable in the data, PIE data is only investigated for time\_based models. \cref{tab:samsize} presents the number of sample sizes generated by the 47 instances of mid-block crosses extracted from PIE dataset. As it can be seen in this table, the number of instances that can be used for training the model over PIE dataset is significantly smaller than the number of instances collected using VR. This reinforces the idea that VR enables the collection of larger amounts of data under specific scenarios. Although for general purposes, video data might give us faster and more realistic choices for data collection. 
\begin{table}[!h]
\centering
\footnotesize
\caption{Number of samples used for training and testing of PIE dataset}
\begin{tabular}{|ll|}
\hline
\textbf{Time-based} & \textbf{Samples} \\ \hline
   $T\_1\_1$               & 2,051            \\ \hline
   $T\_1\_2$               & 1,581            \\ \hline
  $T\_2\_1$               & 1,581            \\ \hline
\end{tabular}
\label{tab:samsize}
\end{table}

\cref{tab:pieresults} shows the results of 8-fold cross-validation for the top models trained on PIE dataset. The similar configurations of the best models in the PIE dataset and VR dataset show that the structures found for the VR dataset can be applied to the video dataset without a further need to find the best hyperparameters. Out of the 6 best models in \cref{tab:timemodels}  and \cref{tab:pieresults}, four of them have the exact same configurations, and the difference among the other two is limited to only one parameter. Similar to \cref{tab:timemodels}, it can be seen that adding auxiliary information to the model improves the error of the prediction over validation and training data. The best improvement is observed over T\_1\_2, where the movements of the pedestrian in the next two second is predicted with the prior 1 seconds of movements. This is also in line with the results of the VR dataset, where auxiliary information is most helpful when the amount of prior time-series data is the least.

\begin{table}[!h]
\centering
\caption{8-fold cross-validation results for top time-based models trained of PIE dataset. Errors are reported in pixels as root mean square error of the center of the bounding boxes over all predicted time steps}
\label{tab:pieresults}
\scalebox{0.68}{
\begin{tabular}{|llllllllllll|}
\hline
\textbf{$t_1$ }    & \textbf{$t_2$ }    & \textbf{Type} & \textbf{\begin{tabular}[c]{@{}l@{}}Dense\\ Layers\end{tabular}} & \textbf{\begin{tabular}[c]{@{}l@{}}LSTM \\ Layers\end{tabular}} & \textbf{Nodes} & \textbf{\begin{tabular}[c]{@{}l@{}}Batch \\ Size\end{tabular}} & \textbf{Dropout} & \textbf{\begin{tabular}[c]{@{}l@{}}Validation \\ Loss \end{tabular}} & \multicolumn{1}{l}{\textbf{\begin{tabular}[c]{@{}l@{}}Validation \\ Error \end{tabular}}} & \textbf{\begin{tabular}[c]{@{}l@{}}Train \\ Loss \end{tabular}} & \textbf{\begin{tabular}[c]{@{}l@{}}Train \\ Error \end{tabular}} \\ \hline
\multirow{2}{*}{1} & \multirow{2}{*}{1} & Aux-LSTM      & 3                                                               & 2                                                               & 100            & 32                                                             & 0                & 943                                                                    & 31                                                                                        & 1233                                                               & 35                                                             \\
                   &                    & Vanilla       & NA                                                              & 2                                                               & 100            & 32                                                             & 0                & 1451                                                                   & 38                                                                                        & 2249                                                               & 47                                                              \\ \hline
\multirow{2}{*}{1} & \multirow{2}{*}{2} & Aux-LSTM      & 3                                                               & 2                                                               & 100            & 32                                                             & 0                & 2031                                                                   & 45                                                                                        & 1995                                                               & 44                                                              \\
                   &                    & Vanilla       & NA                                                              & 2                                                               & 100            & 32                                                             & 0                & 6312                                                                    & 79                                                                                        & 9070                                                               & 95                                                              \\ \hline
\multirow{2}{*}{2} & \multirow{2}{*}{1} & Aux-LSTM      & 1                                                               & 3                                                               & 100            & 32                                                             & 0                & 907                                                                    & 30                                                                                        & 725                                                               & 27                                                              \\
                   &                    & Vanilla       & NA                                                              & 2                                                               & 100            & 32                                                             & 0                & 2031                                                                    & 45                                                                                        & 1523                                                               & 39                                                              \\ \hline
\end{tabular}}
\end{table}

\subsection{Interpretation of Results}
\begin{figure}[!h]
    \centering
    \includegraphics[scale=0.6]{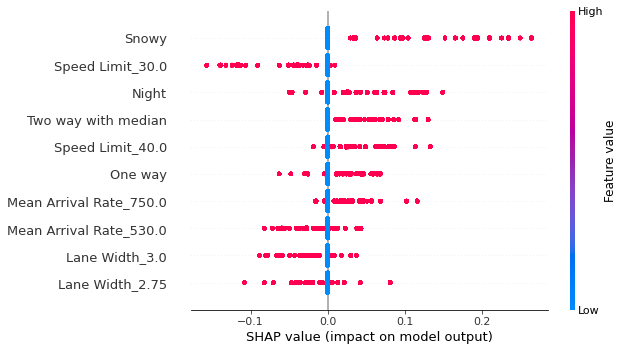}
    \caption{Plot summary of the effects of auxiliary variables on error}
    \label{fig:Tshap}
\end{figure}

\arash{The results of applying SHAP to a selected trained model are presented in this section.} \cref{fig:Tshap} plots the summary of the SHAP values applied to the Aux-LSTM model trained on T\_1\_1 data. The effects of variables on the error of predicting pedestrian trajectory are provided in this plot. Having a positive SHAP value for a variable in an instance means higher error due to the presence of that factor.

The most contributing variable to the error is snow. According to the figure, in all the instances with snowy weather, the prediction error is increased. A similar trend holds in night scenarios, with the majority of instances in night scenarios leading to an increase in prediction error. With the affected sight distance in the night and snowy environments, participants were expected to follow more erratic trajectories. The high impact of weather conditions in the VR environment shows the importance of having environmental diversity in AV datasets. Regarding the variable related to speed limit, it can be observed in the plot that at lower speeds, the models can predict the trajectory more accurately. It can be concluded that in our experiment, participants were behaving more predictably when confronting slower traffic. The same behaviour can be seen in scenarios with lower vehicle flow rates. In most instances with a low flow rate (530 veh/hr), the SHAP value of the corresponding arrival rate variable is negative, meaning the positive impact of this variable on achieving higher accuracy in trajectory prediction. Assuming more stress levels of participants when confronting faster or more congested traffics, more uncertain and unpredictable trajectories in these conditions can be explained. Another significantly contributing variable seems to be the road type. In the scenarios with a two-way road and median, the prediction has higher errors, showing a more unpredictable trajectory of pedestrians when facing vehicles in two directions. Finally, lane width variables do not seem to have a consistent impact on the instances, with SHAP values in different instances spreading to both sides of the spectrum. In general, contextual information on traffic characteristics, road geometry, and weather conditions appear to have the greatest impact on the error of the model. Based on these observations, we recommend considering such variables during the data collection and modelling pedestrian behaviour when interacting with AVs. All the variables included in the model were set so that a typical AV can capture or calculate based on the information obtained by its camera, LiDAR, or other sensors available to them.

\section{Conclusions and Future Works}
\label{S:t6}
Pedestrian trajectory prediction models can be used in various automated contexts, e.g., automated vehicles or automated delivery robots. By having a better estimation of pedestrian's future behaviour based on their current behaviour, we can ensure a safe and comfortable trip for both pedestrians and passengers in the vehicle. \arash{Moreover, accurate prediction of pedestrian trajectory leads to a more efficient choice of speed for the vehicles, as well as the minimization of unnecessary breaks and stops, meaning smoother traffic flows on urban roads.}

In this study, we explored the use of naturalistic virtual reality data and advanced machine learning models to predict pedestrian crossing trajectories. In the proposed method, contextual information from the environment is used as auxiliary data. The auxiliary data are then added to sequential data of pedestrian's past trajectory, head orientations and distance to the upcoming vehicles, to train an LSTM network for predicting pedestrians' next coordinates. By adding auxiliary data, our framework takes into account the effects of road specifications, i.e., lane width and type of road, traffic parameters, i.e., speed limit, arrival rate, and environmental conditions, i.e. weather conditions and time of the day. All the auxiliary variables are chosen in a way that a hypothetical AV can observe and use the information for its prediction algorithm. 

To show the generalizability of the proposed model, we applied the proposed methodology to sensors data of pedestrian trajectories, extracted from PIE dataset. The results showed that incorporating contextual information within the trajectory prediction models increases the prediction accuracy, on both VR and video data. By implementing a neural network interpretability method, we conclude that a pedestrian-oriented AV dataset requires to include diverse weather and vision conditions, as well as different traffic conditions, to be able to predict and model pedestrian trajectories accurately. \arash{Despite the growing accessibility of open-access AV datasets, a major part of the currently available datasets fails to provide such variety in environmental conditions. Furthermore, currently available open-access AV datasets often lack adequate information of pedestrians on specific crossing conditions.} AV manufacturers can use our methodological framework and results to better understand the contextual factors that can negatively affect their prediction algorithms and try to address the possible shortcomings by changing the focal point of their data collection efforts to include problematic situations. \arash{Providers of open-access datasets to this point have mainly focused on improvements of object detection, annotations and vehicle-oriented tasks, and a lesser amount of focus has been dedicated to pedestrians. Collecting and publishing datasets that focus on particular types of interactions, e.g. with pedestrians or cyclists, can help research communities to develop more accurate and generalizable models, ultimately leading to a safer urban area. Furthermore, controlled data collections can be utilized to include a wider range of demographics who might not be represented in data collections concluded in particular areas.}  

This study is not without limitations, which are remained to be explored in future studies. Although pedestrian intention and waiting time can be determining factors in predicting the next movements of pedestrians, the current study does not account for this effect. A joint model consisting of both the pedestrian intention and trajectory can provide a comprehensive tool for AVs to predict the behaviour of pedestrians in a broader sense. Although we demonstrated an application of our framework on real-world video data, utilizing models from other state-of-the-art trajectory prediction methods, and transferring the ideas behind them to the context of interest in this study can be another way to compare the performance of our methodology to other studies. Moreover, comparing the trajectory of pedestrians when facing AVs and regular vehicles can help understand the expected changes in the future urban areas.
In the future steps of the study, the two-way communication and training of the AV can also be explored. Redefining the problem to include the vehicle side of the interaction with pedestrians considering the comfort and safety of passengers is another possible dimension to discover in future studies. One of the objectives of this study was to introduce features that can be used to improve the prediction accuracy of trajectory models. With a collaboration with computer vision experts, extracting these features from an available AV dataset and applying the model trained on VR data to the AV datasets would lead to a better understanding of the capabilities of VR data. Finally, presenting a hybrid dataset on pedestrian behaviours when confronting AVs by incorporating various open datasets as well as VR data can be a direction to follow in future studies. Training on a hybrid dataset would allow a more generalized model, which can benefit from the collective advantages of different data sources.

\bibliographystyle{elsarticle-harv}

\newpage
\appendix
\section{Extended cross-validation results for trajectory prediction models}

\label{A:valT}
The results of the best configurations of Aux-LSTM and vanilla LSTM models trained using different combinations of the time-series VR data as input are provided in this section. The best configurations are chosen based on 8-fold cross-validation results. Time-series data derived from the VR reality include pedestrian coordinates, head orientations, and distance to the ego vehicle. To find out the impact of the addition of head orientation and distance to vehicle on predicting the coordinates of pedestrians, four variants of the models were developed, where each variant receives a subset of time-series data as input. These variants and the cross-validation results for their top-performing models are presented here. In all the following tables, errors are reported in meters as root mean square error over all predicted time steps.\\
\textit{1. Variant-xy:} receives solely pedestrians' coordinates ($x_0,y_0$) as time-series input:
\begin{table}[!h]
\centering
\caption{8-fold cross-validation results for top distance-based models trained on VR data: Variant-xy}
\label{tab:distmodelsxy}
\scalebox{0.7}{
\begin{tabular}{|lllllllllll|}
\hline
\textbf{p}           & \textbf{Type} & \textbf{\begin{tabular}[c]{@{}l@{}}Dense\\ Layers\end{tabular}} & \textbf{\begin{tabular}[c]{@{}l@{}}LSTM \\ Layers\end{tabular}} & \textbf{Nodes} & \textbf{\begin{tabular}[c]{@{}l@{}}Batch \\ Size\end{tabular}} & \textbf{Dropout} & \textbf{\begin{tabular}[c]{@{}l@{}}Validation \\ Loss \end{tabular}} & \multicolumn{1}{l}{\textbf{\begin{tabular}[c]{@{}l@{}}Validation \\ Error \end{tabular}}} & \textbf{\begin{tabular}[c]{@{}l@{}}Train \\ Loss \end{tabular}} & \textbf{\begin{tabular}[c]{@{}l@{}}Train \\ Error \end{tabular}} \\ \hline
\multirow{2}{*}{0.3} & Aux-LSTM      & 2                                                               & 2                                                               & 100            & 32                                                             & 0                &   0.0354 & 0.1882                            &   0.0374	&0.1935                                                              \\
                     & Vanilla       & NA                                                              & 1                                                               & 100            & 32                                                             & 0                & 0.0299&	0.1729&	0.0348&	0.1866                                                         \\ \hline
\multirow{2}{*}{0.5} & Aux-LSTM      & 2                                                               & 2                                                               & 50             & 128                                                            & 0                &     0.0230                                                                & 0.1518                                                                                         & 0.0190                                                                &    0.1382                                                           \\
                     & Vanilla       & NA                                                              & 1                                                               & 50             & 32                                                            & 0                & 0.0183 &	0.1354 &	0.0163	& 0.1278                                                              \\ \hline
\multirow{2}{*}{0.7} & Aux-LSTM      & 1                                                               & 3                                                               & 50             & 32                                                             & 0                &  0.0053                                                                   & 0.0727                                                                                         & 0.0047                                                                &   0.0692                                                            \\
                     & Vanilla       & NA                                                              & 1                                                               & 50             & 32                                                             & 0                & 0.0042&	0.0644&	0.0038&	0.0619                                                              \\ \hline
\end{tabular}}
\end{table}
\begin{table}[!h]
\centering
\caption{8-fold cross-validation results for top time-based models trained on VR data: Variant-xy}
\label{tab:timemodelsxy}
\scalebox{0.68}{
\begin{tabular}{|llllllllllll|}
\hline
\textbf{$t_1$ }    & \textbf{$t_2$ }    & \textbf{Type} & \textbf{\begin{tabular}[c]{@{}l@{}}Dense\\ Layers\end{tabular}} & \textbf{\begin{tabular}[c]{@{}l@{}}LSTM \\ Layers\end{tabular}} & \textbf{Nodes} & \textbf{\begin{tabular}[c]{@{}l@{}}Batch \\ Size\end{tabular}} & \textbf{Dropout} & \textbf{\begin{tabular}[c]{@{}l@{}}Validation \\ Loss \end{tabular}} & \multicolumn{1}{l}{\textbf{\begin{tabular}[c]{@{}l@{}}Validation \\ Error \end{tabular}}} & \textbf{\begin{tabular}[c]{@{}l@{}}Train \\ Loss \end{tabular}} & \textbf{\begin{tabular}[c]{@{}l@{}}Train \\ Error \end{tabular}} \\ \hline
\multirow{2}{*}{1} & \multirow{2}{*}{1} & Aux-LSTM      & 3                                                               & 2                                                               & 100            & 32                                                             & 0                &    0.0263                                                                 &    0.1622                                                                                     & 0.0215&	0.1466                                                                                                                            \\
                   &                    & Vanilla       & NA                                                              & 1                                                               & 100            & 32                                                             & 0                & 0.0519&	0.2279&	0.0507&	0.2251                                                           \\ \hline
\multirow{2}{*}{1} & \multirow{2}{*}{2} & Aux-LSTM      & 3                                                               & 2                                                               & 100            & 32                                                             & 0                &    0.0459                                                                 &    0.2142                                                                                     & 0.0427&	0.2065                                                                                                                             \\
                   &                    & Vanilla       & NA                                                              & 2                                                               & 100            & 32                                                             & 0                & 0.1883&	0.4340&	0.1758&	0.4193                                                           \\ \hline
\multirow{2}{*}{2} & \multirow{2}{*}{1} & Aux-LSTM      & 3                                                               & 3                                                               & 100            & 32                                                             & 0                &                                                                  0.0150   &    0.1225                                                                                     &          0.0137	&0.1172                                                             \\
                   &                    & Vanilla       & NA                                                              & 3                                                               & 100            & 32                                                             & 0                & 0.0437&	0.2090&	0.0385&	0.1963                                                           \\ \hline
\end{tabular}}
\end{table}
\\
\textit{2. Variant-xyo:} receives pedestrians' coordinates ($x_0,y_0$) and head orientations ($o_0$) as time-series input:
\begin{table}[!h]
\centering
\caption{8-fold cross-validation results for top distance-based models trained on VR data: Variant-xyo}
\label{tab:distmodelsxyo}
\scalebox{0.7}{
\begin{tabular}{|lllllllllll|}
\hline
\textbf{p}           & \textbf{Type} & \textbf{\begin{tabular}[c]{@{}l@{}}Dense\\ Layers\end{tabular}} & \textbf{\begin{tabular}[c]{@{}l@{}}LSTM \\ Layers\end{tabular}} & \textbf{Nodes} & \textbf{\begin{tabular}[c]{@{}l@{}}Batch \\ Size\end{tabular}} & \textbf{Dropout} & \textbf{\begin{tabular}[c]{@{}l@{}}Validation \\ Loss \end{tabular}} & \multicolumn{1}{l}{\textbf{\begin{tabular}[c]{@{}l@{}}Validation \\ Error \end{tabular}}} & \textbf{\begin{tabular}[c]{@{}l@{}}Train \\ Loss \end{tabular}} & \textbf{\begin{tabular}[c]{@{}l@{}}Train \\ Error \end{tabular}} \\ \hline
\multirow{2}{*}{0.3} & Aux-LSTM      & 2                                                               & 3                                                               & 100            & 32                                                             & 0                & 0.0258	&0.1605	&0.0172	&0.1312                                                        \\
                     & Vanilla       & NA                                                              & 1                                                               & 100            & 32                                                             & 0                & 0.0205&	0.1431&	0.0130&	0.1139                                                              \\ \hline
\multirow{2}{*}{0.5} & Aux-LSTM      & 3                                                               & 2                                                               & 50             & 128                                                            & 0                & 0.0112&	0.1059&	0.0095&	0.0972                                                       \\
                     & Vanilla       & NA                                                              & 1                                                               & 100             & 32                                                            & 0                & 0.0071&	0.0840&	0.0066&	0.0812                                                              \\ \hline
\multirow{2}{*}{0.7} & Aux-LSTM      & 1                                                               & 3                                                               & 50             & 32                                                             & 0                &0.0051                                                                     & 0.0711                                                                                        & 0.0047                                                                &                      0.0682                                         \\
                     & Vanilla       & NA                                                              & 3                                                               & 100             & 32                                                             & 0                & 0.0035                & 	0.0589                & 	0.0017	                & 0.0409                                                              \\ \hline
\end{tabular}}
\end{table}
\begin{table}[!h]
\centering
\caption{8-fold cross-validation results for top time-based models trained on VR data: Variant-xyo}
\label{tab:timemodelsxyo}
\scalebox{0.68}{
\begin{tabular}{|llllllllllll|}
\hline
\textbf{$t_1$ }    & \textbf{$t_2$ }    & \textbf{Type} & \textbf{\begin{tabular}[c]{@{}l@{}}Dense\\ Layers\end{tabular}} & \textbf{\begin{tabular}[c]{@{}l@{}}LSTM \\ Layers\end{tabular}} & \textbf{Nodes} & \textbf{\begin{tabular}[c]{@{}l@{}}Batch \\ Size\end{tabular}} & \textbf{Dropout} & \textbf{\begin{tabular}[c]{@{}l@{}}Validation \\ Loss \end{tabular}} & \multicolumn{1}{l}{\textbf{\begin{tabular}[c]{@{}l@{}}Validation \\ Error \end{tabular}}} & \textbf{\begin{tabular}[c]{@{}l@{}}Train \\ Loss \end{tabular}} & \textbf{\begin{tabular}[c]{@{}l@{}}Train \\ Error \end{tabular}} \\ \hline
\multirow{2}{*}{1} & \multirow{2}{*}{1} & Aux-LSTM      & 2                                                               & 3                                                               & 100            & 32                                                             & 0                & 0.0239&	0.1547&	0.0115&	0.1070                                                        \\
                   &                    & Vanilla       & NA                                                              & 3                                                               & 100            & 32                                                             & 0                & 0.0378&	0.1944&	0.0230&	0.1516                                                           \\ \hline
\multirow{2}{*}{1} & \multirow{2}{*}{2} & Aux-LSTM      & 2                                                               & 2                                                               & 100            & 32                                                             & 0                & 0.0518&	0.2275&	0.0216&	0.1468                                                        \\
                   &                    & Vanilla       & NA                                                              & 3                                                               & 100            & 32                                                             & 0                & 0.0958&	0.3095&	0.0447&	0.2115                                                           \\ \hline
\multirow{2}{*}{2} & \multirow{2}{*}{1} & Aux-LSTM      & 2                                                               & 3                                                               & 100            & 32                                                             & 0                & 0.0182&	0.1351&	0.0126&	0.1124                                                        \\
                   &                    & Vanilla       & NA                                                              & 3                                                               & 100            & 32                                                             & 0                & 0.0107&	0.1032&	0.0063&	0.0793                                                           \\ \hline
\end{tabular}}
\end{table}
\newpage
\noindent \textit{3. Variant-xyd:} receives pedestrians' coordinates ($x_0,y_0$) and distance to vehicles ($d_0$) as time-series input:
\begin{table}[!h]
\centering
\caption{8-fold cross-validation results for top distance-based models trained on VR data: Variant-xyd}
\label{tab:distmodelsxyd}
\scalebox{0.7}{
\begin{tabular}{|lllllllllll|}
\hline
\textbf{p}           & \textbf{Type} & \textbf{\begin{tabular}[c]{@{}l@{}}Dense\\ Layers\end{tabular}} & \textbf{\begin{tabular}[c]{@{}l@{}}LSTM \\ Layers\end{tabular}} & \textbf{Nodes} & \textbf{\begin{tabular}[c]{@{}l@{}}Batch \\ Size\end{tabular}} & \textbf{Dropout} & \textbf{\begin{tabular}[c]{@{}l@{}}Validation \\ Loss \end{tabular}} & \multicolumn{1}{l}{\textbf{\begin{tabular}[c]{@{}l@{}}Validation \\ Error \end{tabular}}} & \textbf{\begin{tabular}[c]{@{}l@{}}Train \\ Loss \end{tabular}} & \textbf{\begin{tabular}[c]{@{}l@{}}Train \\ Error \end{tabular}} \\ \hline
\multirow{2}{*}{0.3} & Aux-LSTM      & 1                                                               & 3                                                               & 100            & 32                                                             & 0                & 0.0255&	0.1598 &	0.0206&	0.1434                                                      \\
                     & Vanilla       & NA                                                              & 3                                                               & 100            & 32                                                             & 0                & 0.0251&	0.1583&	0.0170&	0.1305                                            \\ \hline
\multirow{2}{*}{0.5} & Aux-LSTM      & 1                                                               & 1                                                               & 100             & 32                                                            & 0                & 0.0084&	0.0914&	0.0077&	0.0877                                                        \\
                     & Vanilla       & NA                                                              & 1                                                               & 100             & 32                                                            & 0                & 0.0068&	0.0826&	0.0051&	0.0714                                                              \\ \hline
\multirow{2}{*}{0.7} & Aux-LSTM      & 1                                                               & 3                                                               & 50             & 32                                                             & 0                &    0.0027                                                                 & 0.0521                                                                                       & 0.0024                                                                &                     0.0488                                          \\
                     & Vanilla       & NA                                                              & 2                                                               & 100             & 32                                                             & 0                & 0.0018&	0.0422&	0.0028&	0.0527                                                              \\ \hline
\end{tabular}}
\end{table}
\begin{table}[!h]
\centering
\caption{8-fold cross-validation results for top time-based models trained on VR data: Variant-xyd}
\label{tab:timemodelsxyd}
\scalebox{0.68}{
\begin{tabular}{|llllllllllll|}
\hline
\textbf{$t_1$ }    & \textbf{$t_2$ }    & \textbf{Type} & \textbf{\begin{tabular}[c]{@{}l@{}}Dense\\ Layers\end{tabular}} & \textbf{\begin{tabular}[c]{@{}l@{}}LSTM \\ Layers\end{tabular}} & \textbf{Nodes} & \textbf{\begin{tabular}[c]{@{}l@{}}Batch \\ Size\end{tabular}} & \textbf{Dropout} & \textbf{\begin{tabular}[c]{@{}l@{}}Validation \\ Loss \end{tabular}} & \multicolumn{1}{l}{\textbf{\begin{tabular}[c]{@{}l@{}}Validation \\ Error \end{tabular}}} & \textbf{\begin{tabular}[c]{@{}l@{}}Train \\ Loss \end{tabular}} & \textbf{\begin{tabular}[c]{@{}l@{}}Train \\ Error \end{tabular}} \\ \hline
\multirow{2}{*}{1} & \multirow{2}{*}{1} & Aux-LSTM      & 3                                                               & 2                                                               & 100            & 32                                                             & 0                & 0.0337&	0.1834&	0.0189&	0.1375                                                        \\
                   &                    & Vanilla       & NA                                                              & 2                                                               & 100            & 32                                                             & 0                & 0.0449&	0.2120&	0.0427&	0.2066                                                           \\ \hline
\multirow{2}{*}{1} & \multirow{2}{*}{2} & Aux-LSTM      & 3                                                               & 2                                                               & 100            & 32                                                             & 0                & 0.0651&	0.2551	&0.0326&	0.1804                                                        \\
                   &                    & Vanilla       & NA                                                              & 3                                                               & 100            & 32                                                             & 0                & 0.1542&	0.3927&	0.1261&	0.3552                                                           \\ \hline
\multirow{2}{*}{2} & \multirow{2}{*}{1} & Aux-LSTM      & 3                                                               & 3                                                               & 100            & 32                                                             & 0                & 0.0225&	0.1499&	0.0198&	0.1408                                                        \\
                   &                    & Vanilla       & NA                                                              & 3                                                               & 100            & 32                                                             & 0                & 0.0340 &	0.1844	&0.0319&	0.1787                                                           \\ \hline
\end{tabular}}
\end{table}
\\
\textit{4. Variant-xyod:} receives pedestrians' coordinates ($x_0,y_0$), head orientations ($o_0$) and distance to vehicles ($d_0$) as time-series input:
\begin{table}[!h]
\centering
\caption{8-fold cross-validation results for top distance-based models trained on VR data: Variant-xyod}
\label{tab:distmodelsxyod}
\scalebox{0.7}{
\begin{tabular}{|lllllllllll|}
\hline
\textbf{p}           & \textbf{Type} & \textbf{\begin{tabular}[c]{@{}l@{}}Dense\\ Layers\end{tabular}} & \textbf{\begin{tabular}[c]{@{}l@{}}LSTM \\ Layers\end{tabular}} & \textbf{Nodes} & \textbf{\begin{tabular}[c]{@{}l@{}}Batch \\ Size\end{tabular}} & \textbf{Dropout} & \textbf{\begin{tabular}[c]{@{}l@{}}Validation \\ Loss \end{tabular}} & \multicolumn{1}{l}{\textbf{\begin{tabular}[c]{@{}l@{}}Validation \\ Error \end{tabular}}} & \textbf{\begin{tabular}[c]{@{}l@{}}Train \\ Loss \end{tabular}} & \textbf{\begin{tabular}[c]{@{}l@{}}Train \\ Error \end{tabular}} \\ \hline
\multirow{2}{*}{0.3} & Aux-LSTM      & 2                                                               & 3                                                               & 100            & 32                                                             & 0                & 0.0219                                                                    & 0.1481                                                                                        & 0.0171                                                               & 0.1306                                                              \\
                     & Vanilla       & NA                                                              & 2                                                               & 100            & 32                                                             & 0                & 0.0180                                                                    & 0.1341                                                                                        & 0.0183                                                               & 0.1351                                                              \\ \hline
\multirow{2}{*}{0.5} & Aux-LSTM      & 2                                                               & 2                                                               & 50             & 128                                                            & 0                & 0.0104                                                                    & 0.1021                                                                                        & 0.0115                                                               & 0.1071                                                              \\
                     & Vanilla       & NA                                                              & 1                                                               & 50             & 128                                                            & 0                & 0.0088                                                                    & 0.0940                                                                                        & 0.0071                                                               & 0.0841                                                              \\ \hline
\multirow{2}{*}{0.7} & Aux-LSTM      & 1                                                               & 3                                                               & 50             & 32                                                             & 0                & 0.0069                                                                    & 0.0830                                                                                        & 0.0059                                                               & 0.0769                                                              \\
                     & Vanilla       & NA                                                              & 1                                                               & 10             & 32                                                             & 0                & 0.0025                                                                    & 0.0500                                                                                        & 0.0030                                                               & 0.0545                                                              \\ \hline
\end{tabular}}
\end{table}
\begin{table}[!h]
\centering
\caption{8-fold cross-validation results for top time-based models trained on VR data: Variant-xyod}
\label{tab:timemodelsxyod}
\scalebox{0.68}{
\begin{tabular}{|llllllllllll|}
\hline
\textbf{$t_1$ }    & \textbf{$t_2$ }    & \textbf{Type} & \textbf{\begin{tabular}[c]{@{}l@{}}Dense\\ Layers\end{tabular}} & \textbf{\begin{tabular}[c]{@{}l@{}}LSTM \\ Layers\end{tabular}} & \textbf{Nodes} & \textbf{\begin{tabular}[c]{@{}l@{}}Batch \\ Size\end{tabular}} & \textbf{Dropout} & \textbf{\begin{tabular}[c]{@{}l@{}}Validation \\ Loss \end{tabular}} & \multicolumn{1}{l}{\textbf{\begin{tabular}[c]{@{}l@{}}Validation \\ Error \end{tabular}}} & \textbf{\begin{tabular}[c]{@{}l@{}}Train \\ Loss \end{tabular}} & \textbf{\begin{tabular}[c]{@{}l@{}}Train \\ Error \end{tabular}} \\ \hline
\multirow{2}{*}{1} & \multirow{2}{*}{1} & Aux-LSTM      & 3                                                               & 2                                                               & 100            & 32                                                             & 0                & 0.0181                                                                    & 0.1344                                                                                        & 0.0085                                                               & 0.0922                                                              \\
                   &                    & Vanilla       & NA                                                              & 2                                                               & 100            & 32                                                             & 0                & 0.0312                                                                    & 0.1767                                                                                        & 0.0170                                                               & 0.1304                                                              \\ \hline
\multirow{2}{*}{1} & \multirow{2}{*}{2} & Aux-LSTM      & 3                                                               & 2                                                               & 100            & 32                                                             & 0                & 0.0305                                                                    & 0.1748                                                                                        & 0.0139                                                               & 0.1178                                                              \\
                   &                    & Vanilla       & NA                                                              & 3                                                               & 100            & 32                                                             & 0                & 0.0781                                                                    & 0.2795                                                                                        & 0.0322                                                               & 0.1795                                                              \\ \hline
\multirow{2}{*}{2} & \multirow{2}{*}{1} & Aux-LSTM      & 3                                                               & 3                                                               & 100            & 32                                                             & 0                & 0.0068                                                                    & 0.0827                                                                                        & 0.0049                                                               & 0.0700                                                              \\
                   &                    & Vanilla       & NA                                                              & 2                                                               & 100            & 32                                                             & 0                & 0.0097                                                                    & 0.0983                                                                                        & 0.0053                                                               & 0.0726                                                              \\ \hline
\end{tabular}}
\end{table}

\newpage Cross-validation results provided in the tables above reveal that in general, the accuracy in prediction obtained by models in all the variants follows a similar pattern. In time-based models, the addition of auxiliary information helps reduce the validation error, whereas, in distance-based models, vanilla models tend to perform better. Comparing the results of different variants, it can be seen that the addition of both head orientations and distance to vehicle to pedestrian coordinates as input improves the validation accuracy of the models, and the best performance is achieved when they are all incorporated simultaneously (variant-xyod). Comparing other variants, it appears that within time-based models, the addition of head orientation improves the performance of the model more than distance to vehicles. This is particularly more significant in vanilla models, which might be due to the lack of input information in this type of developed models. Among distance-based models, the differences between different variants is more subtle, with distance to vehicle showing to have a slightly better contribution in decreasing the validation error.

\section{Trajectory prediction samples on test data}
\label{A:figures}
\begin{figure}[!h]%
 \centering
 \subfloat[slow walk]{\includegraphics[scale=0.35]{L3_sp.5.pdf}\label{fig:a}} \quad
 \subfloat[regular walk]{\includegraphics[scale=0.35]{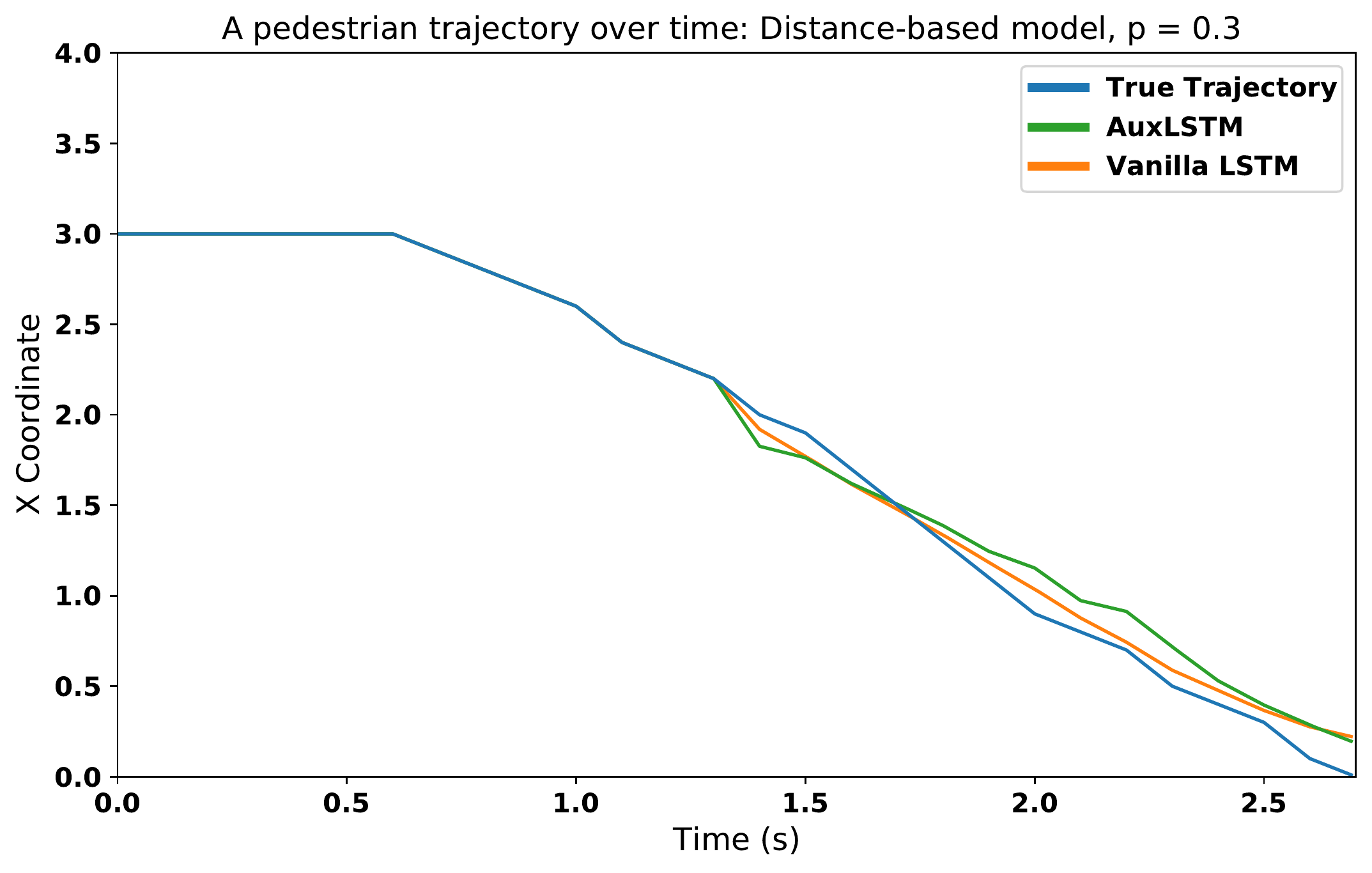}\label{fig:b}}\\
 \subfloat[fast walk]{\includegraphics[scale=0.35]{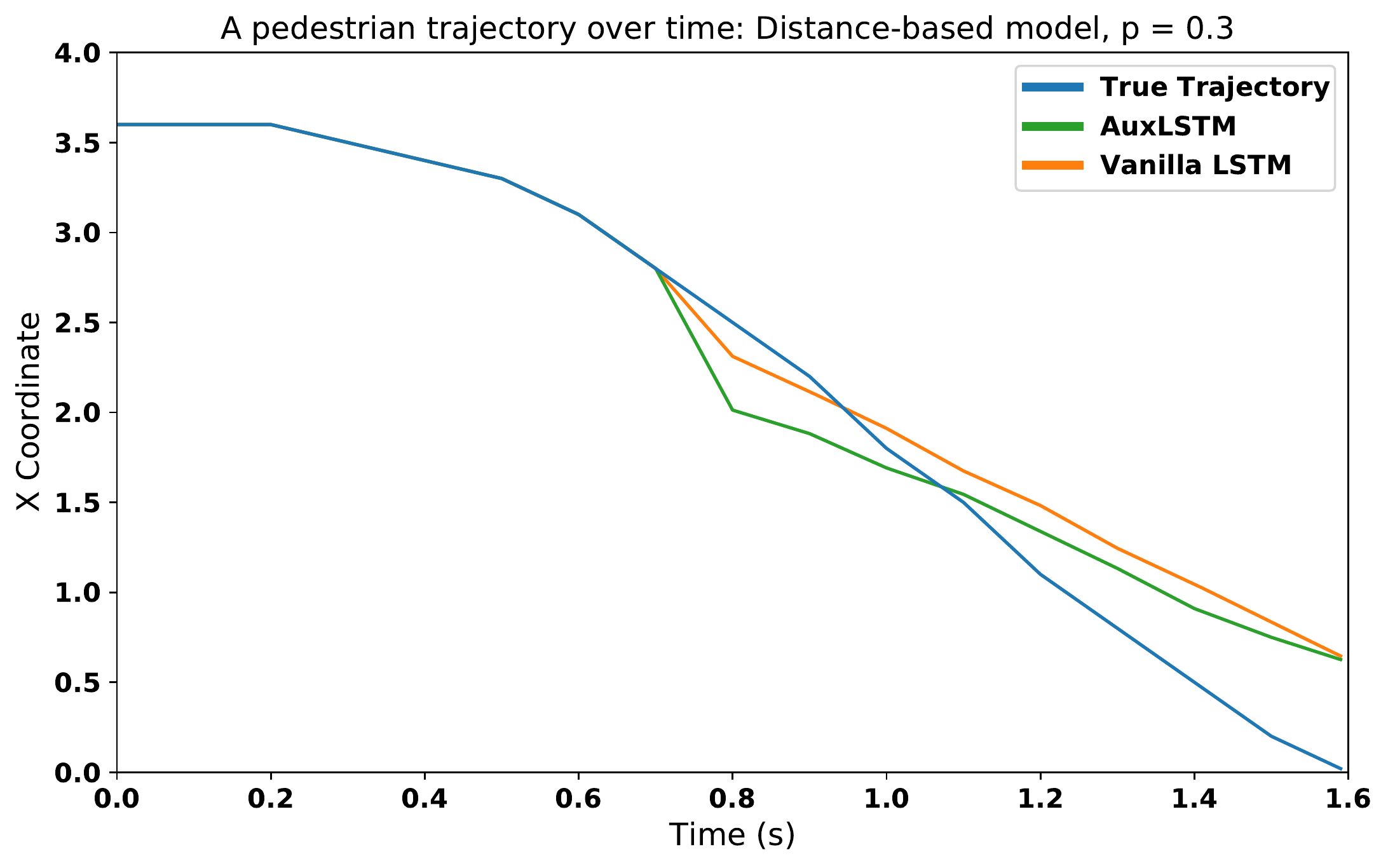}\label{fig:c}}%
 \caption{Distance-based models, p : 0.3}%
 \label{fig:b7}%
\end{figure}

\begin{figure}[!h]%
 \centering
 \subfloat[slow walk]{\includegraphics[scale=0.35]{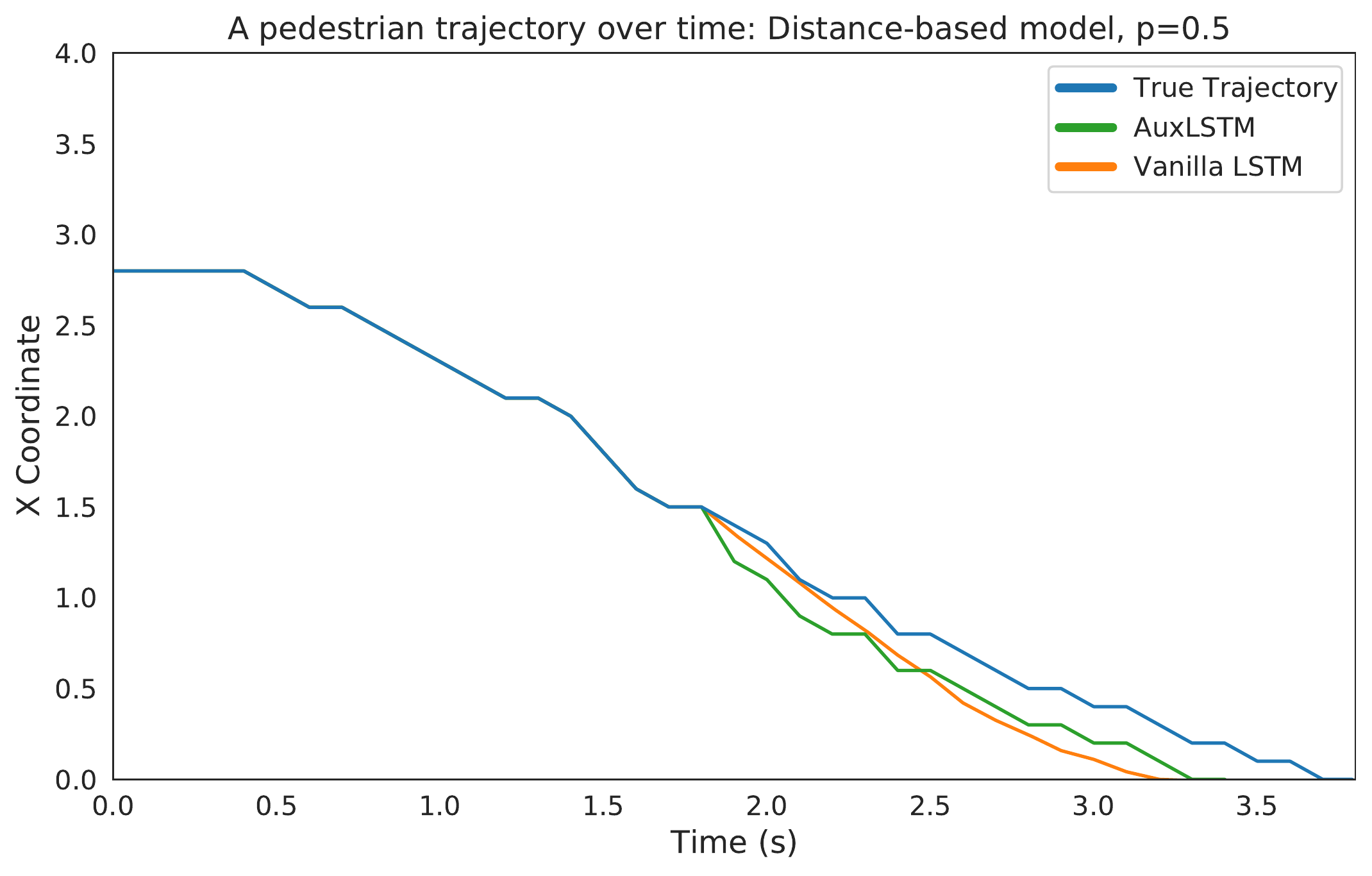}\label{fig:a}} \quad
 \subfloat[regular walk]{\includegraphics[scale=0.35]{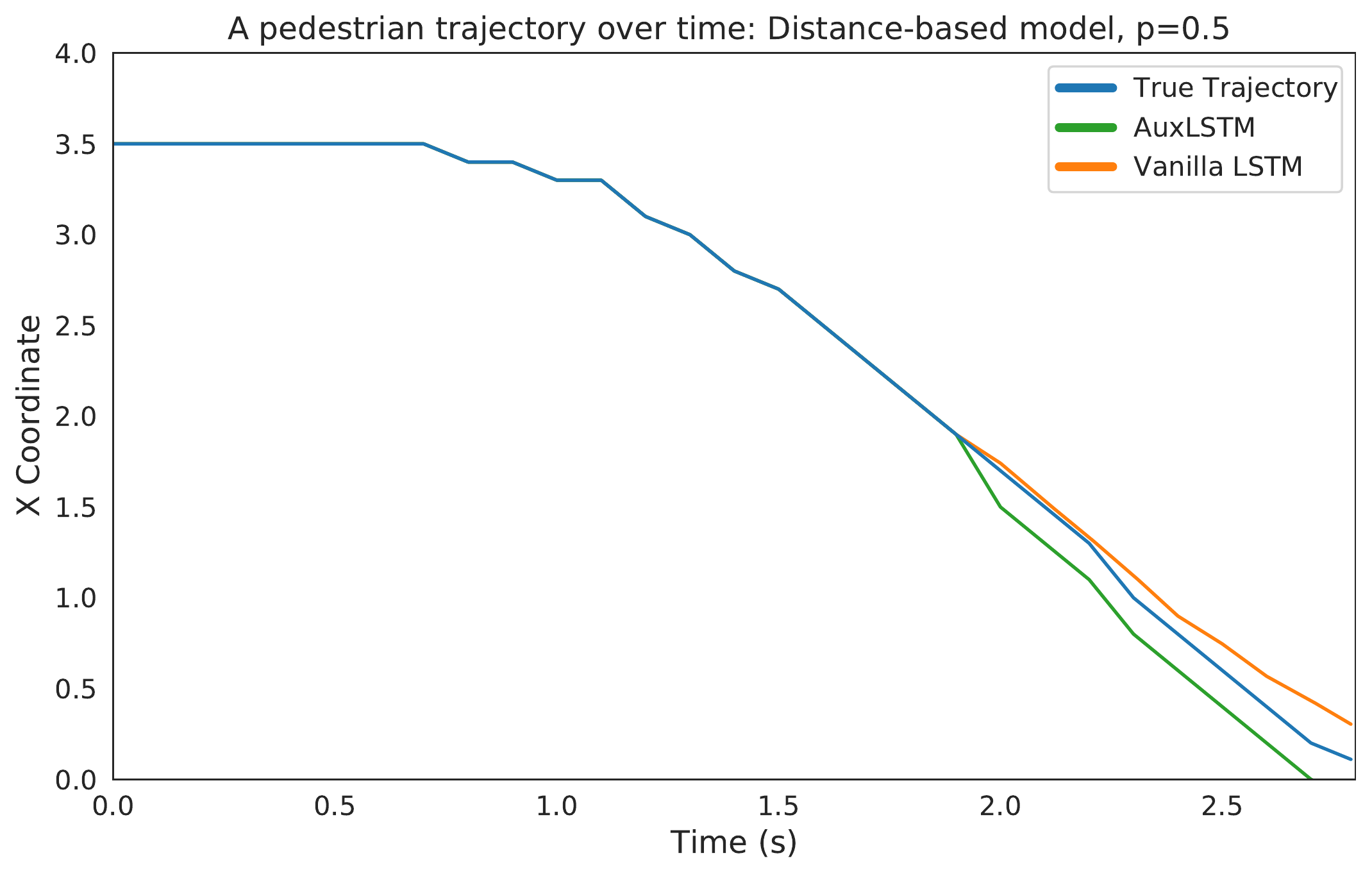}\label{fig:b}}\\
 \subfloat[fast walk]{\includegraphics[scale=0.35]{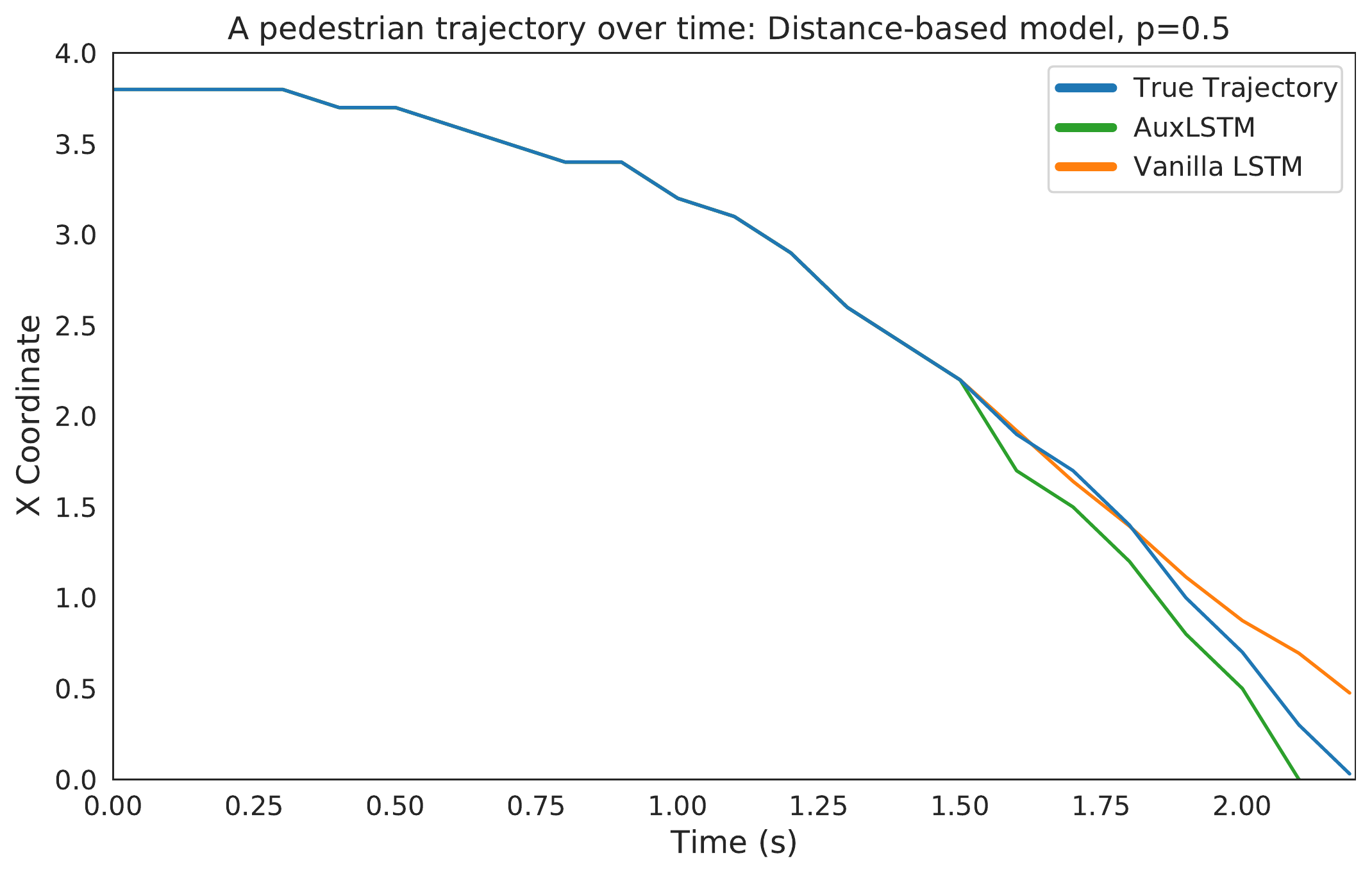}\label{fig:c}}%
 \caption{Distance-based models, p : 0.5}%
 \label{fig:b8}%
\end{figure}

\begin{figure}[!h]%
 \centering
 \subfloat[slow walk]{\includegraphics[scale=0.35]{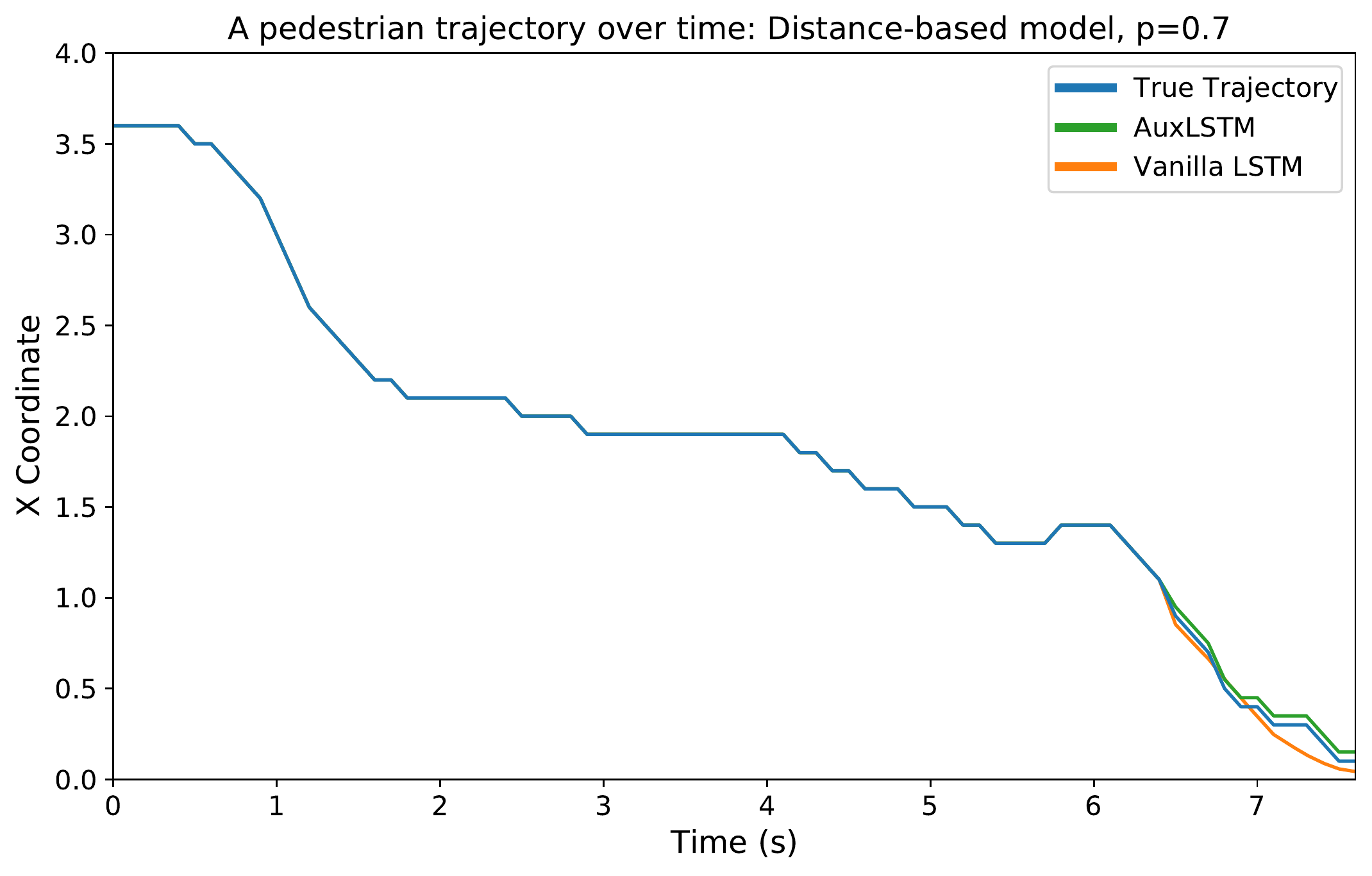}\label{fig:a}} \quad
 \subfloat[regular walk]{\includegraphics[scale=0.35]{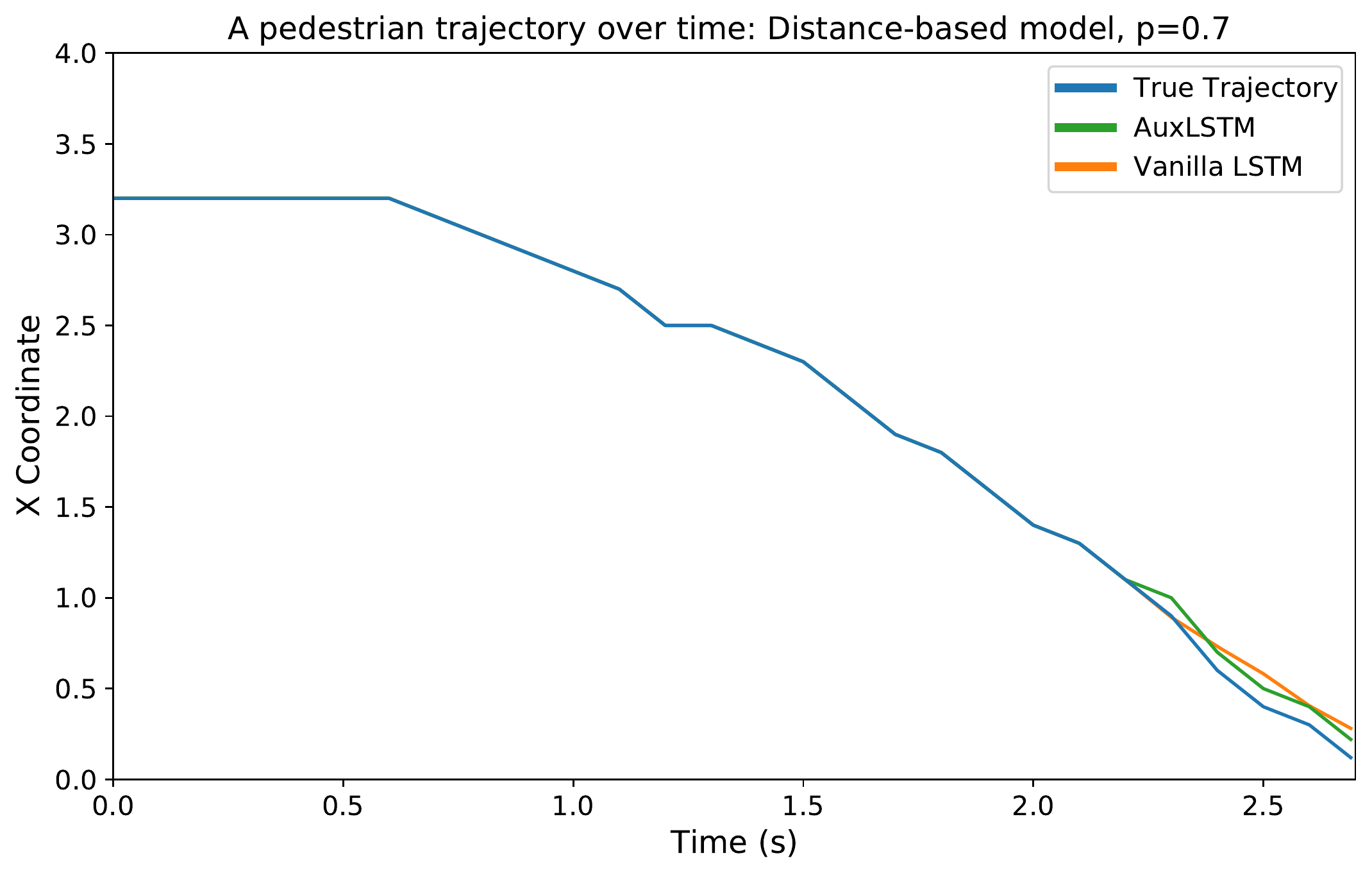}\label{fig:b}}\\
 \subfloat[fast walk]{\includegraphics[scale=0.35]{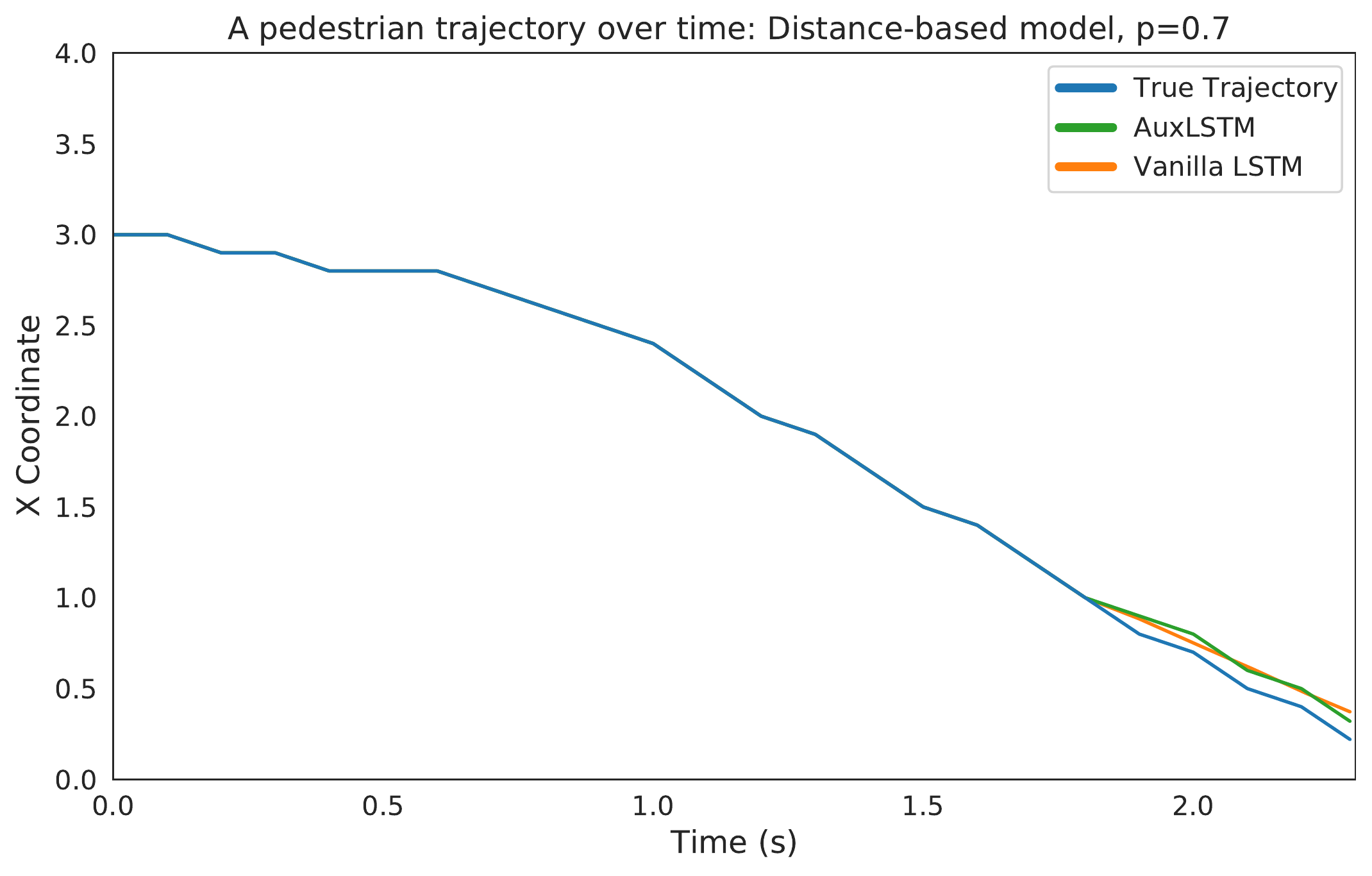}\label{fig:c}}%
 \caption{Distance-based models, p : 0.7}%
 \label{fig:b9}%
\end{figure}

\begin{figure}[!h]%
 \centering
 \subfloat[slow walk]{\includegraphics[scale=0.35]{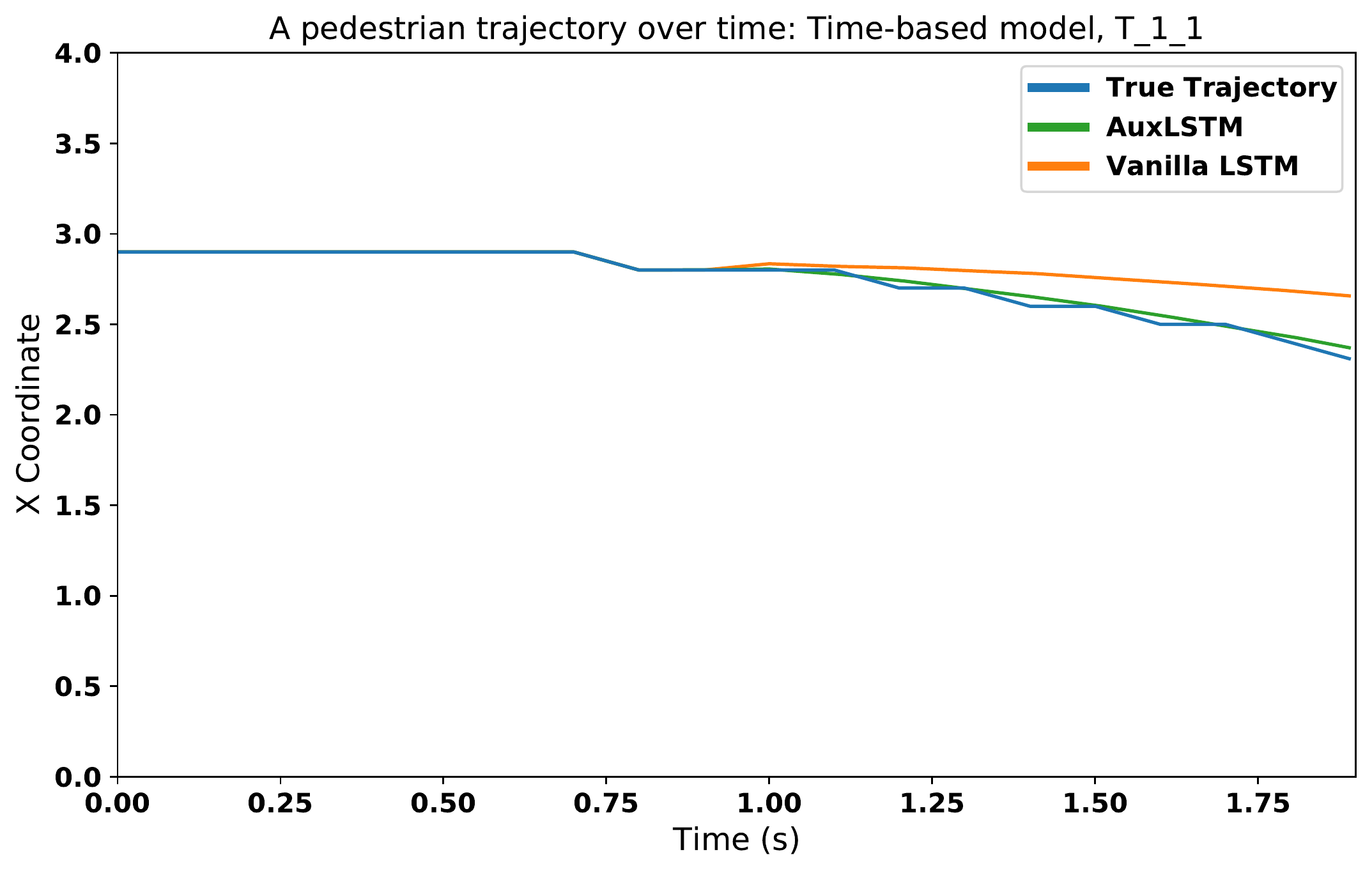}\label{fig:a}} \quad
 \subfloat[regular walk]{\includegraphics[scale=0.35]{T11_sp1.0.pdf}\label{fig:b}}\\
 \subfloat[fast walk]{\includegraphics[scale=0.35]{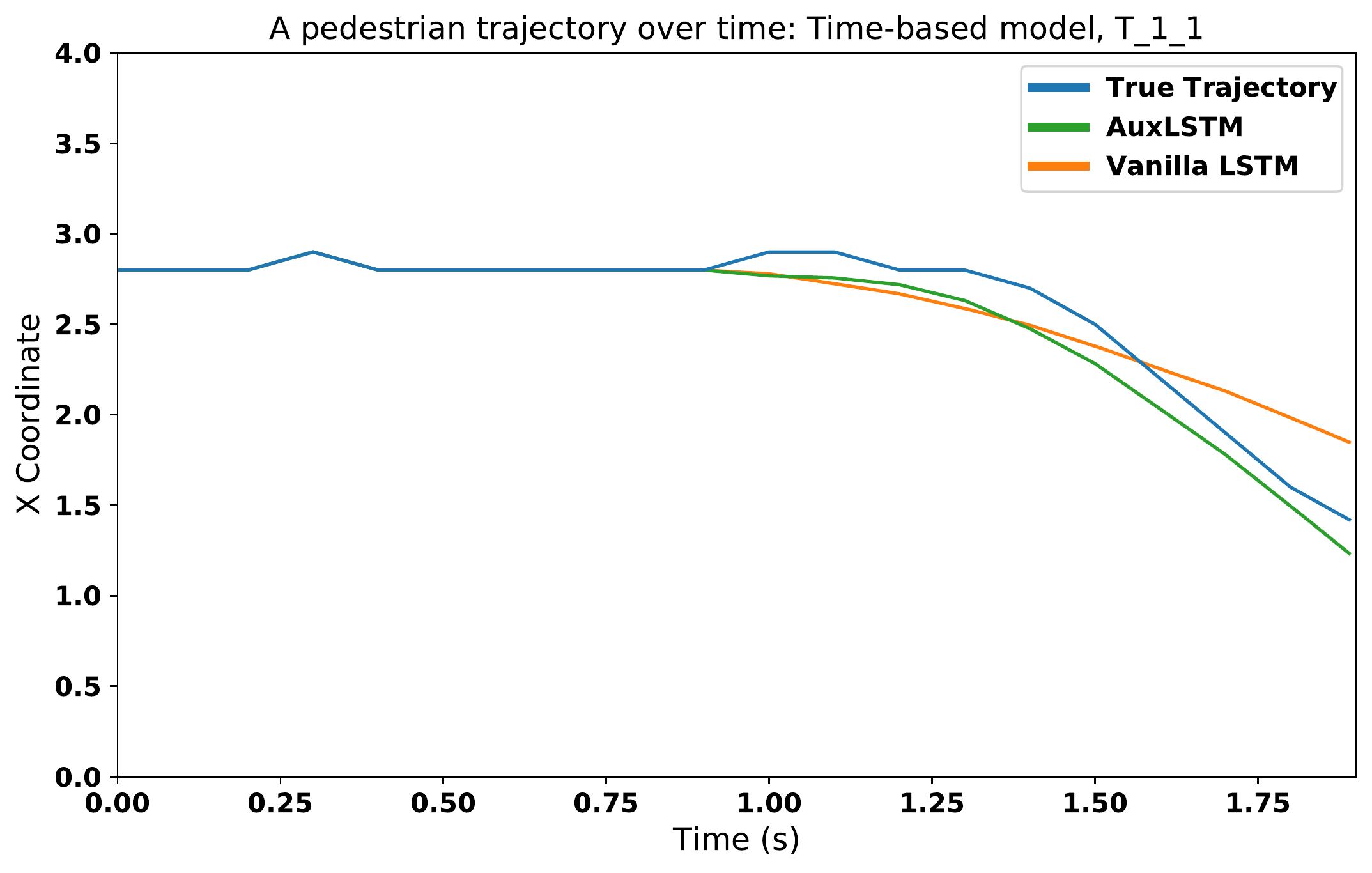}\label{fig:c}}%
 \caption{Time-based models, $t_1$: 1, $t_2$: 1}%
 \label{fig:b10}%
\end{figure}

\begin{figure}[!h]%
 \centering
 \subfloat[slow walk]{\includegraphics[scale=0.35]{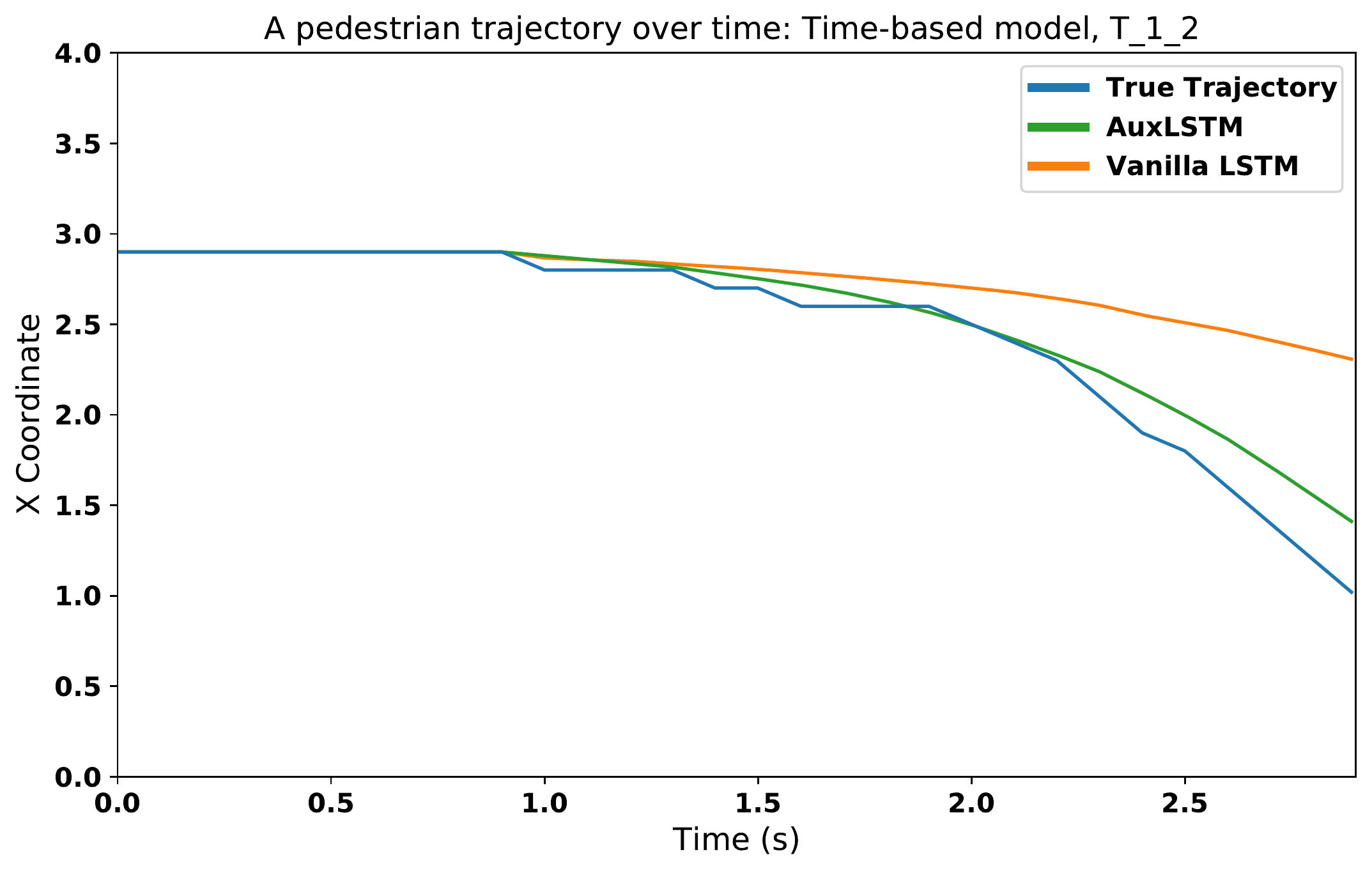}\label{fig:a}} \quad
 \subfloat[regular walk]{\includegraphics[scale=0.35]{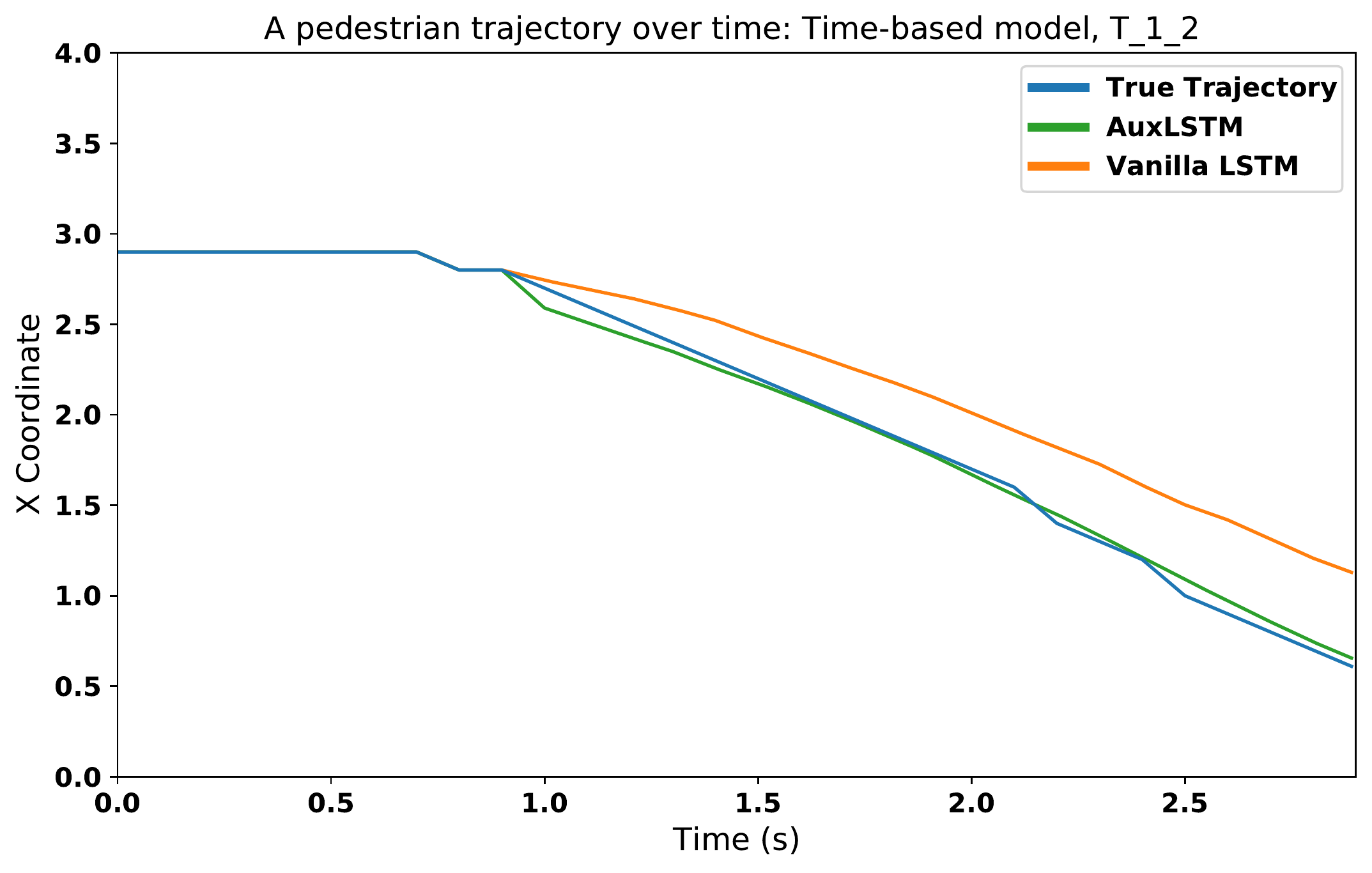}\label{fig:b}}\\
 \subfloat[fast walk]{\includegraphics[scale=0.35]{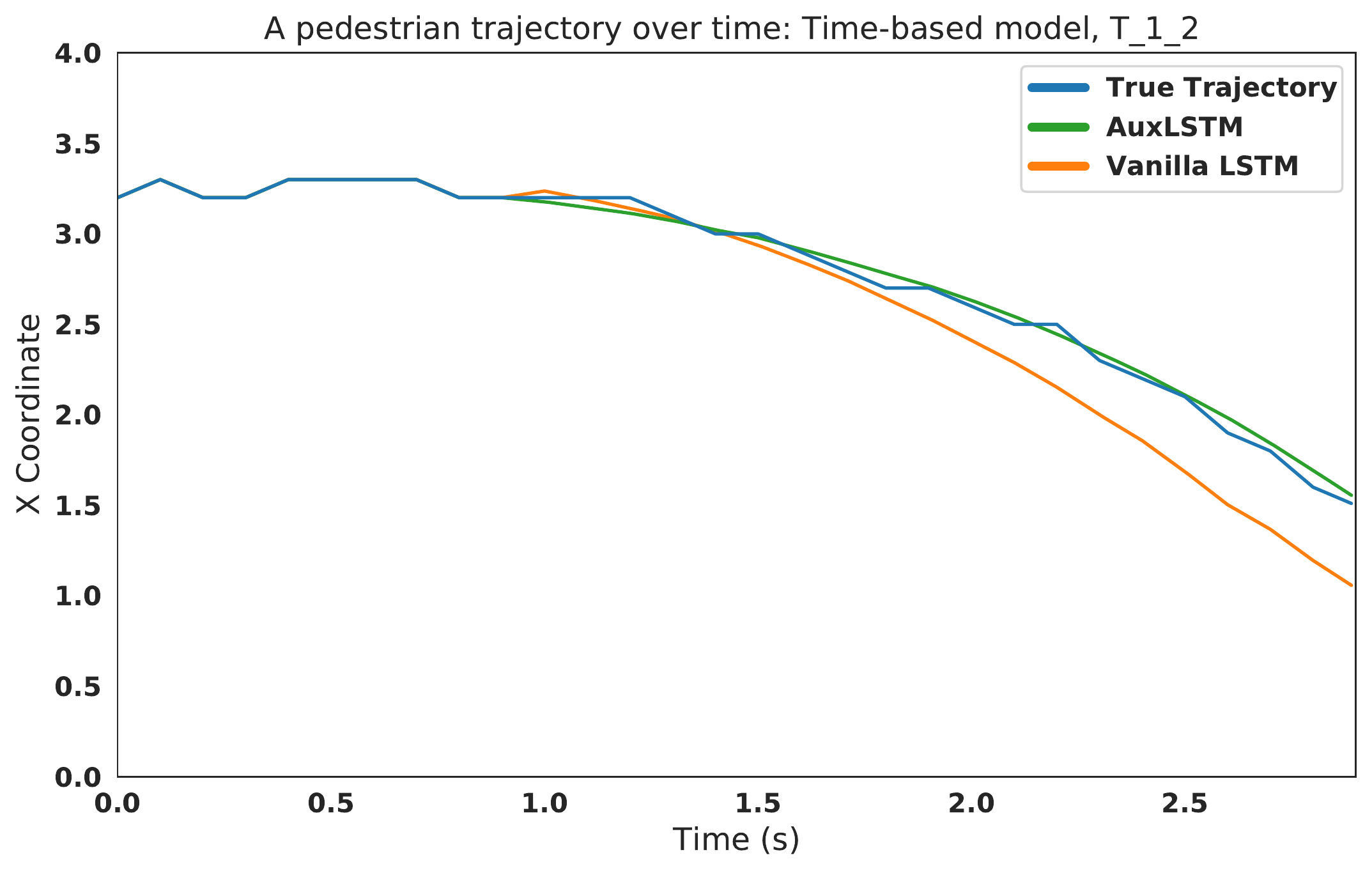}\label{fig:c}}%
 \caption{Time-based models, $t_1$: 1, $t_2$: 2}%
 \label{fig:b11}%
\end{figure}

\begin{figure}[!h]%
 \centering
 \subfloat[slow walk]{\includegraphics[scale=0.35]{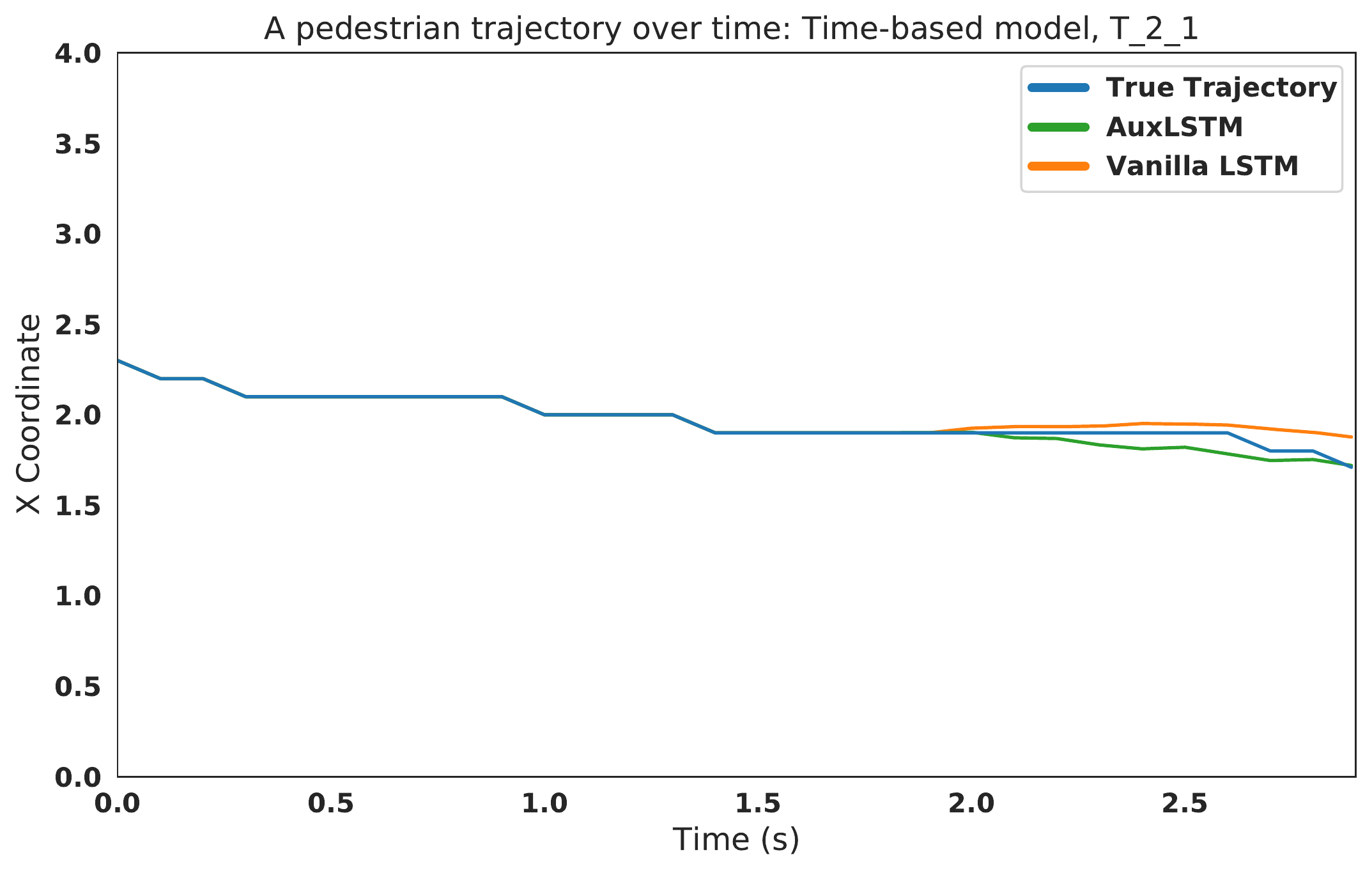}\label{fig:a}} \quad
 \subfloat[regular walk]{\includegraphics[scale=0.35]{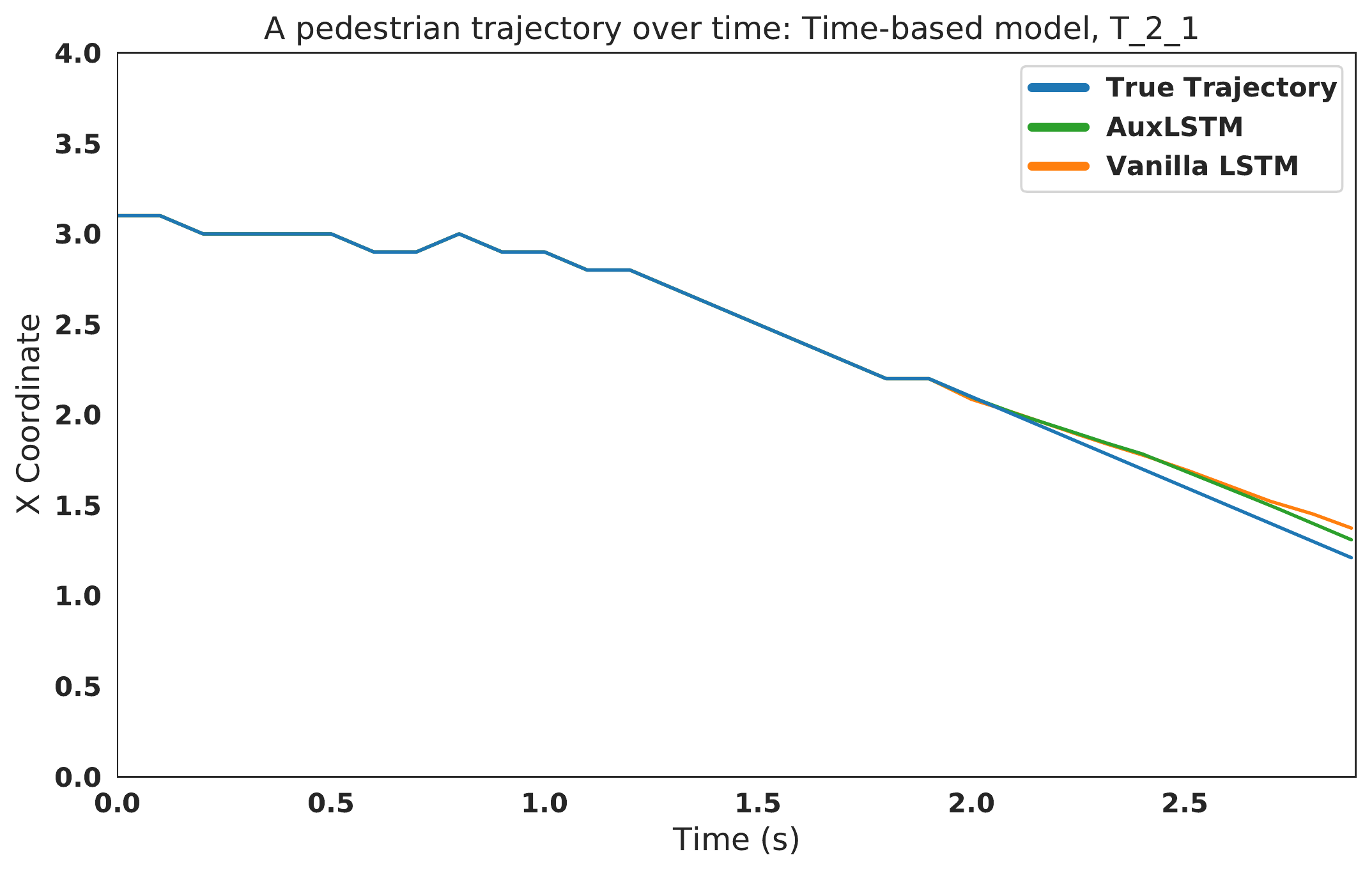}\label{fig:b}}\\
 \subfloat[fast walk]{\includegraphics[scale=0.35]{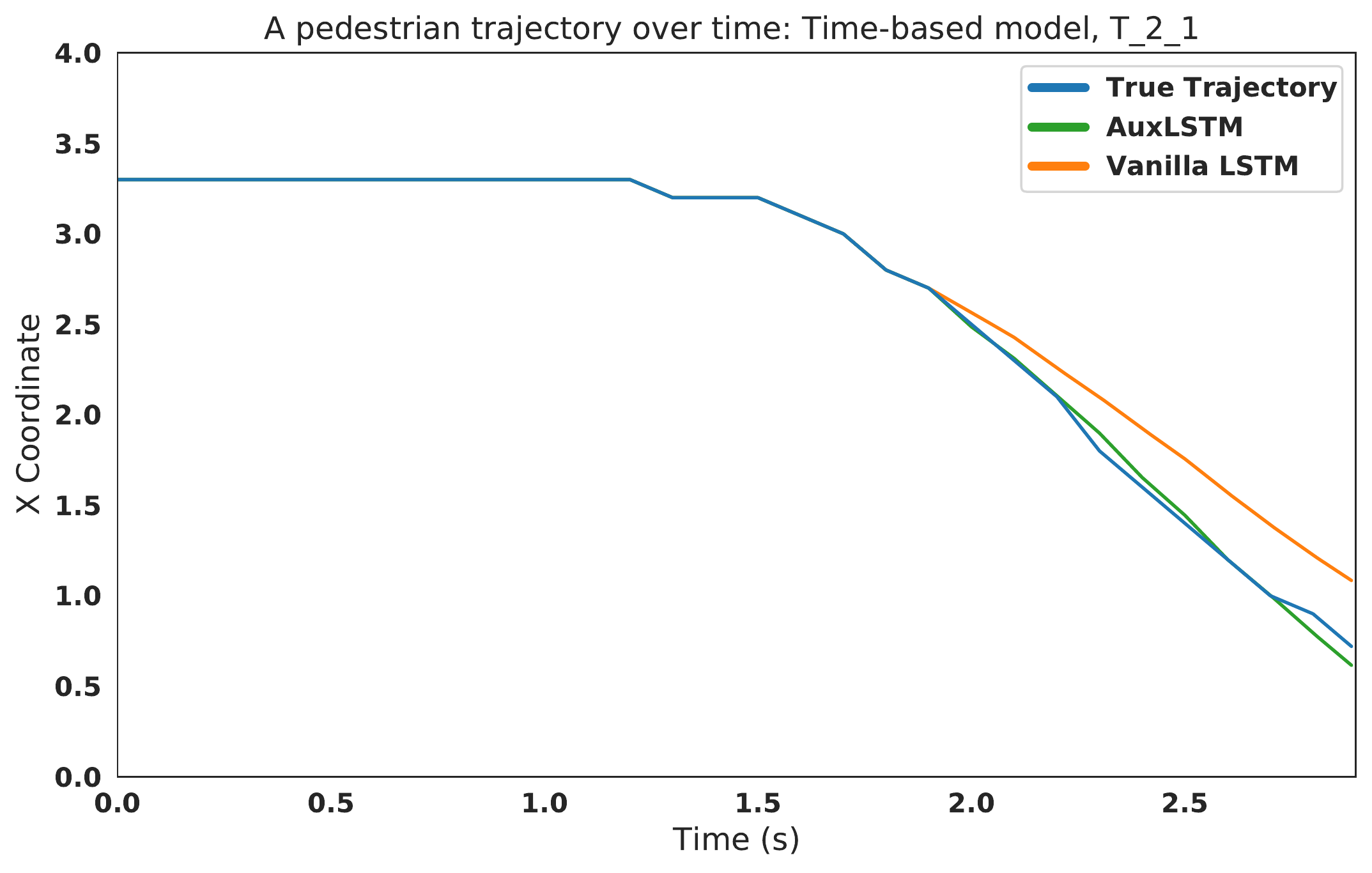}\label{fig:c}}%
 \caption{Time-based models, $t_1$: 2, $t_2$: 1}%
 \label{fig:b12}%
\end{figure}

For each data type, three samples from the test set are selected to depict the prediction performance of the models. For each sample, the ground truth trajectory, along with the predicted trajectory using vanilla and Aux-LSTM models are provided in this section. To show the prediction performance of the models under different conditions, samples include pedestrians with different speeds. It can be observed that the performance of the two modelling approaches over distance-based models varies among the samples, with the fast walks appearing to have the least accurate prediction performance (\cref{fig:b7} to \cref{fig:b9}). On the other hand, the samples confirm the prediction accuracies obtained in \cref{S:t5} that Aux-LSTM outperforms Vanilla LSTM within time-based data (\cref{fig:b10} to \cref{fig:b12}). \\



\end{document}